\documentclass[twocolumn,epjc3]{svjour3}  

\RequirePackage[T1]{fontenc}

\smartqed  

\RequirePackage{graphicx}
\RequirePackage{mathptmx}      
\RequirePackage{flushend}
\RequirePackage[numbers,sort&compress]{natbib}
\RequirePackage[colorlinks,citecolor=blue,urlcolor=blue,linkcolor=blue]{hyperref}

\usepackage{graphics}
\usepackage{amsmath}
\usepackage{subcaption}
\usepackage{siunitx}
\usepackage{xspace}
\usepackage{multirow}
\usepackage{comment}
\usepackage{soul}

\usepackage[switch]{lineno}
\usepackage{etoolbox}

\usepackage[numbers]{natbib}
\smartqed 

\newcommand{\nuBB}{$\beta\beta 2\nu$}
\newcommand{\lessnuBB}{$\beta\beta 0\nu$}
\DeclareSIUnit{\torr}{Torr}

\newcommand{\Bi}[1]{\ensuremath{^{#1}\mathrm{Bi}}\xspace}

\newcommand{\Tl}[1]{\ensuremath{^{#1}\mathrm{Tl}}\xspace}
\newcommand{\Th}[1]{\ensuremath{^{#1}\mathrm{Th}}\xspace}
\newcommand{\U}[1]{\ensuremath{^{#1}\mathrm{U}}\xspace}

\journalname{Eur. Phys. J. C}

\begin{document}

\title{The NEXT-100 Detector}
\author{C. Adams\thanksref{f2,addr1} 
\and
 H. Almaz\'an\thanksref{addr2} 
\and
 V. \'Alvarez\thanksref{addr3} 
\and
 A.I. Aranburu\thanksref{addr2} 
\and
 L. Arazi\thanksref{addr4} 
\and
 I.J. Arnquist\thanksref{addr5} 
\and
 F. Auria-Luna\thanksref{addr6} 
\and
 S. Ayet\thanksref{addr7} 
\and
 Y. Ayyad\thanksref{addr8} 
\and
 C.D.R. Azevedo\thanksref{addr9}
% ...
}

\thankstext{f2}{Now at NVIDIA.}
\thankstext{f26}{Deceased.}
\thankstext{f35}{Now at Weizmann Institute of Science, Israel.}
\thankstext{f37}{On leave.}
\thankstext{f41}{NEXT Spokesperson.}
\thankstext{f89}{Now at University of North Carolina, USA.}
\thankstext{f90}{Now at Los Alamos National Laboratory, USA.}

\institute{
Argonne National Laboratory, Argonne, IL 60439, USA \label{addr1} 
\and
 Donostia International Physics Center, BERC Basque Excellence Research Centre, Manuel de Lardizabal 4, San Sebasti\'an / Donostia, E-20018, Spain \label{addr2} 
\and
 Instituto de Instrumentaci\'on para Imagen Molecular (I3M), Centro Mixto CSIC - Universitat Polit\`ecnica de Val\`encia, Camino de Vera s/n, Valencia, E-46022, Spain \label{addr3} 
\and
 Unit of Nuclear Engineering, Faculty of Engineering Sciences, Ben-Gurion University of the Negev, P.O.B. 653, Beer-Sheva, 8410501, Israel \label{addr4} 
\and
 Pacific Northwest National Laboratory (PNNL), Richland, WA 99352, USA \label{addr5} 
\and
 Department of Organic Chemistry I, Universidad del Pais Vasco (UPV/EHU), Centro de Innovaci\'on en Qu\'imica Avanzada (ORFEO-CINQA), San Sebasti\'an / Donostia, E-20018, Spain \label{addr6} 
\and
 Instituto de F\'isica Corpuscular (IFIC), CSIC \& Universitat de Val\`encia, Calle Catedr\'atico Jos\'e Beltr\'an, 2, Paterna, E-46980, Spain \label{addr7} 
\and
 Instituto Gallego de F\'isica de Altas Energ\'ias, Univ.\ de Santiago de Compostela, Campus sur, R\'ua Xos\'e Mar\'ia Su\'arez N\'u\~nez, s/n, Santiago de Compostela, E-15782, Spain \label{addr8} 
\and
 Institute of Nanostructures, Nanomodelling and Nanofabrication (i3N), Universidade de Aveiro, Campus de Santiago, Aveiro, 3810-193, Portugal \label{addr9} 
% ...
}

\date{Received: date / Accepted: date}

\maketitle

\begin{abstract}
The NEXT collaboration is dedicated to the study of double beta decays of $^{136}$Xe using a high-pressure gas electroluminescent time projection chamber. This advanced technology combines exceptional energy resolution ($\leq 1\%$ FWHM at the $Q_{\beta\beta}$ value of the neutrinoless double beta decay) and powerful topological event discrimination.
Building on the achievements of the NEXT-White detector, the NEXT-100 detector started taking data at the Laboratorio Subterr\'aneo de Canfranc (LSC) in May of 2024. Designed to operate with xenon gas at \qty{13.5}{\bar}, NEXT-100 consists of a time projection chamber where the energy and the spatial pattern of the ionising particles in the detector are precisely retrieved using two sensor planes (one with photo-multiplier tubes and the other with silicon photo-multipliers). The detector has been operating at stable conditions using argon and xenon gases at $\sim$\qty{4}{bar} and drift fields of \qty{74}{\volt\per\centi\meter} and \qty{118}{\volt\per\centi\meter}, respectively. Alpha decays from the $^{222}$Rn chain have been used to test and monitor the stability of the detector, showing a constant electron lifetime in the drift volume. In this paper, in addition to reporting the results of the commissioning run, we provide a detailed description of the NEXT-100 detector, describe its assembly, and present the current estimation of the radiopurity budget. 
\end{abstract}

\begin{figure*}[t!]
  \centering
    \begin{subfigure}{0.50\textwidth}
        \centering
        \rotatebox{0}{\includegraphics[width=1.0\linewidth]{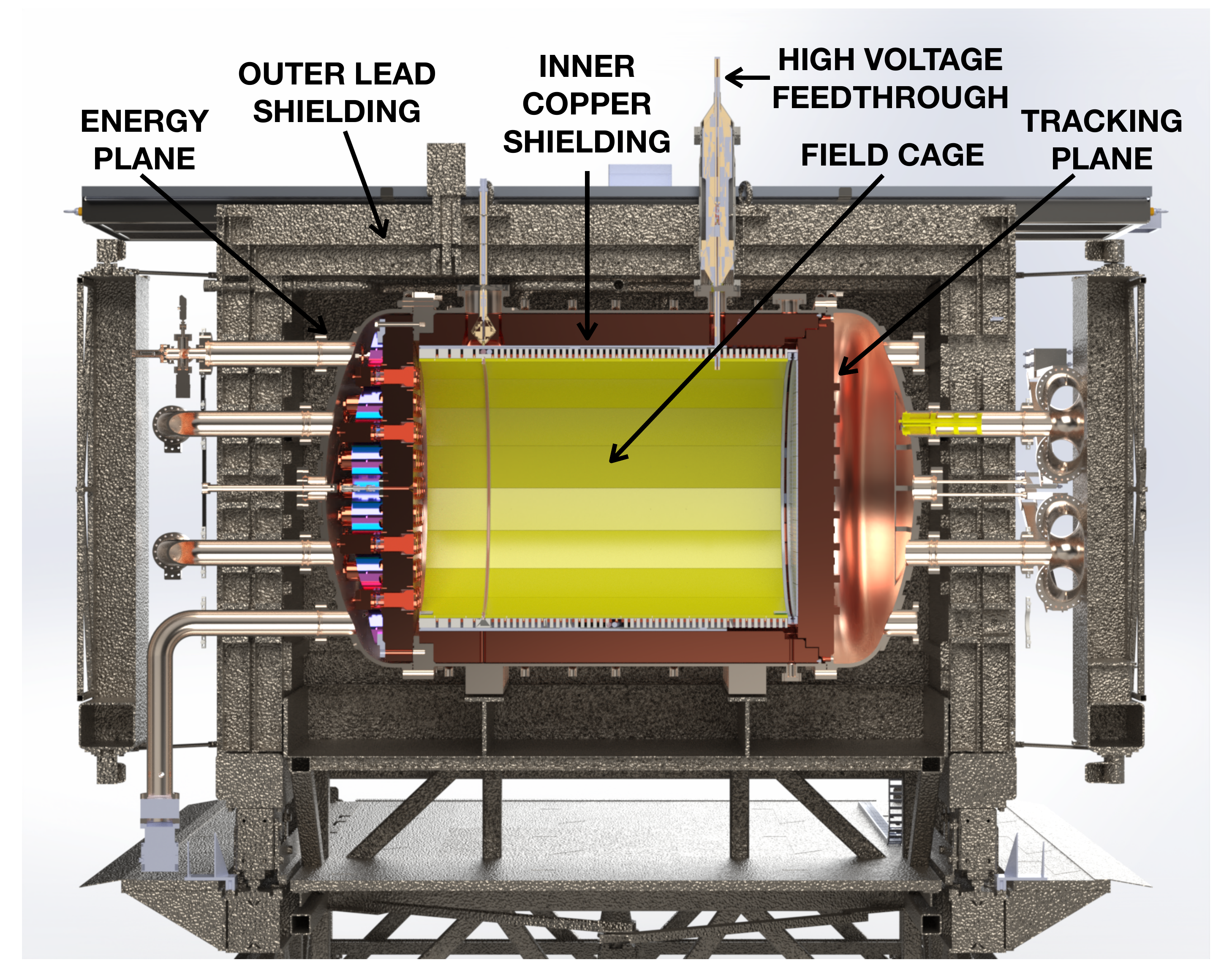}}
        \caption{}
        \label{fig:NEXT100_detector}
    \end{subfigure}
    \hspace{5mm}
    \begin{subfigure}{0.40\textwidth}
        \centering
        \includegraphics[width=1.0\linewidth]{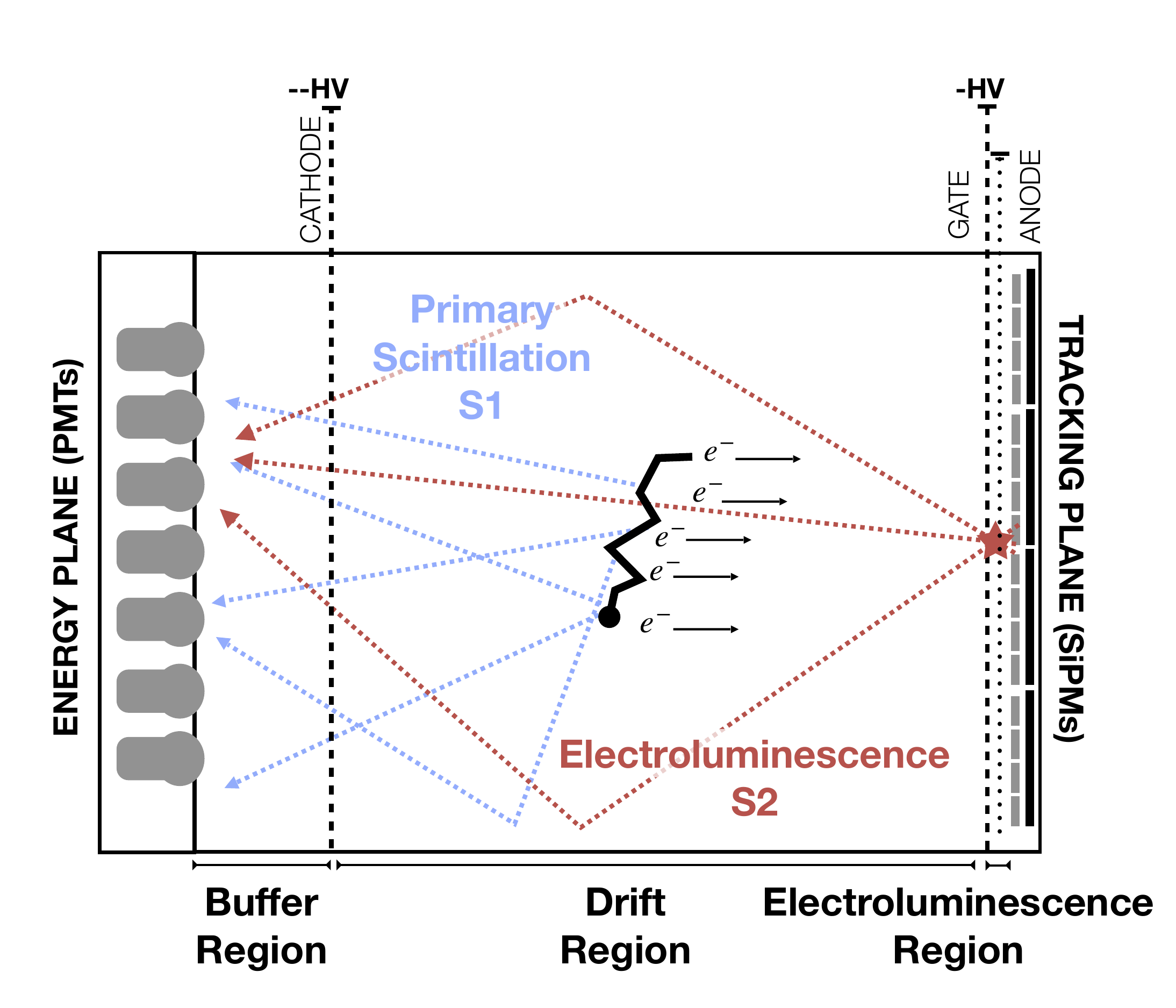}
        \caption{}
        \label{fig:HPXeTPC}
    \end{subfigure}\\
    \begin{subfigure}{0.30\textwidth}
        \centering
        \rotatebox{0}{\includegraphics[width=0.95\linewidth]{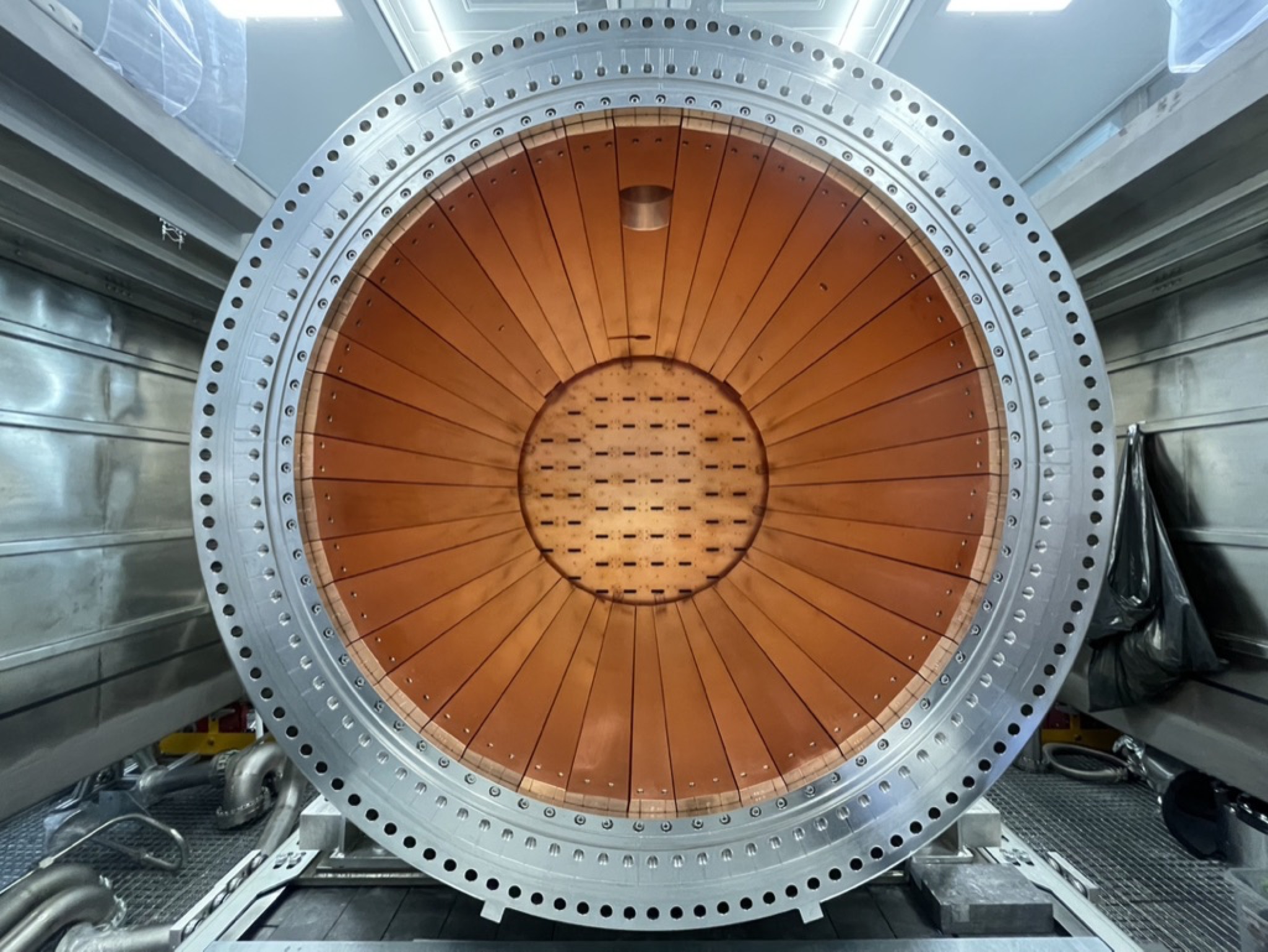}}
        \caption{}
        \label{fig:VesselICS}
    \end{subfigure}
    \begin{subfigure}{0.30\textwidth}
        \centering
        \rotatebox{270}{\includegraphics[width=0.72\linewidth]{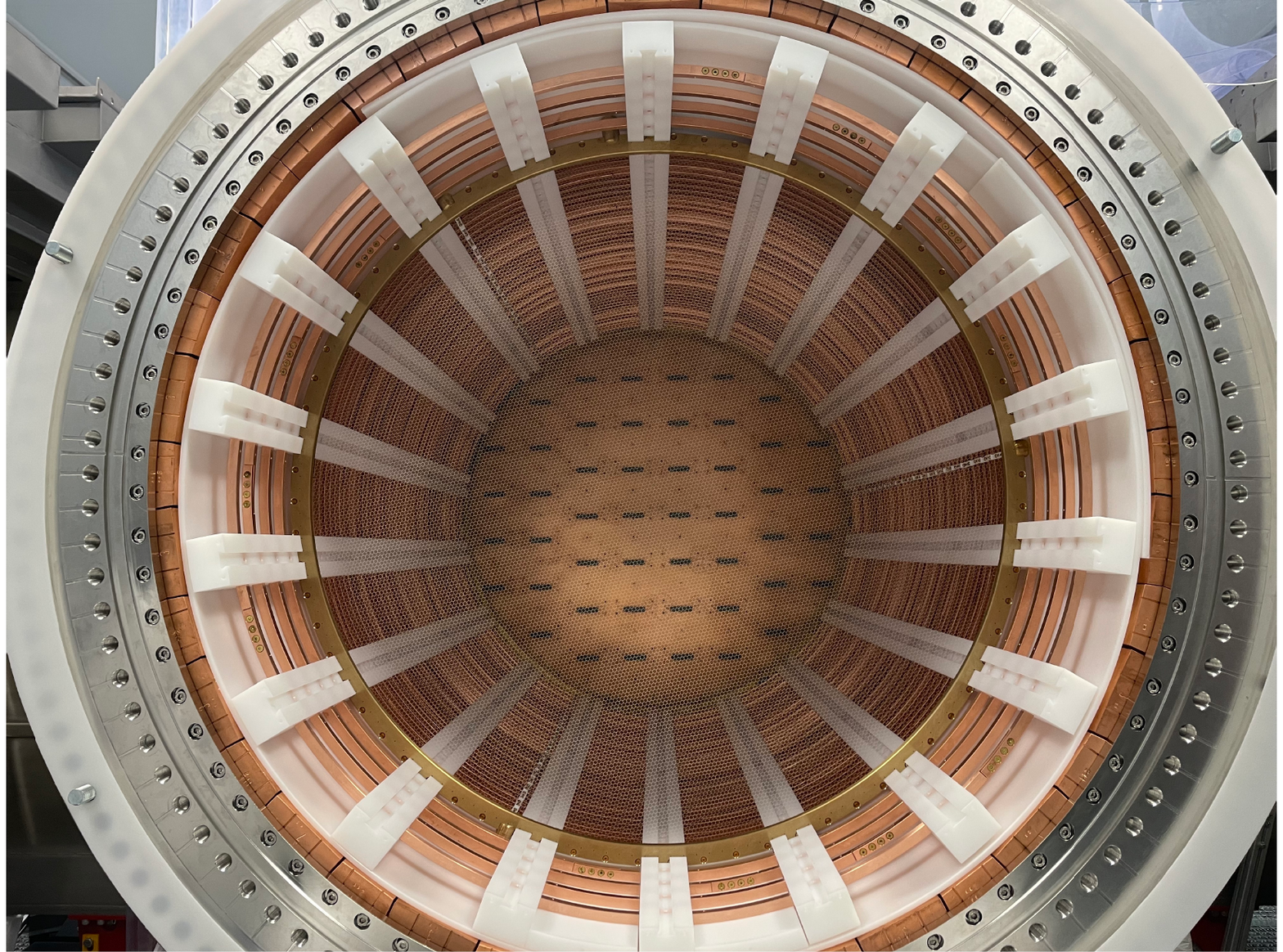}}
        \caption{}
        \label{fig:FC-noteflon}
    \end{subfigure}
    \begin{subfigure}{0.30\textwidth}
        \centering
        \rotatebox{270}{\includegraphics[width=0.72\linewidth]{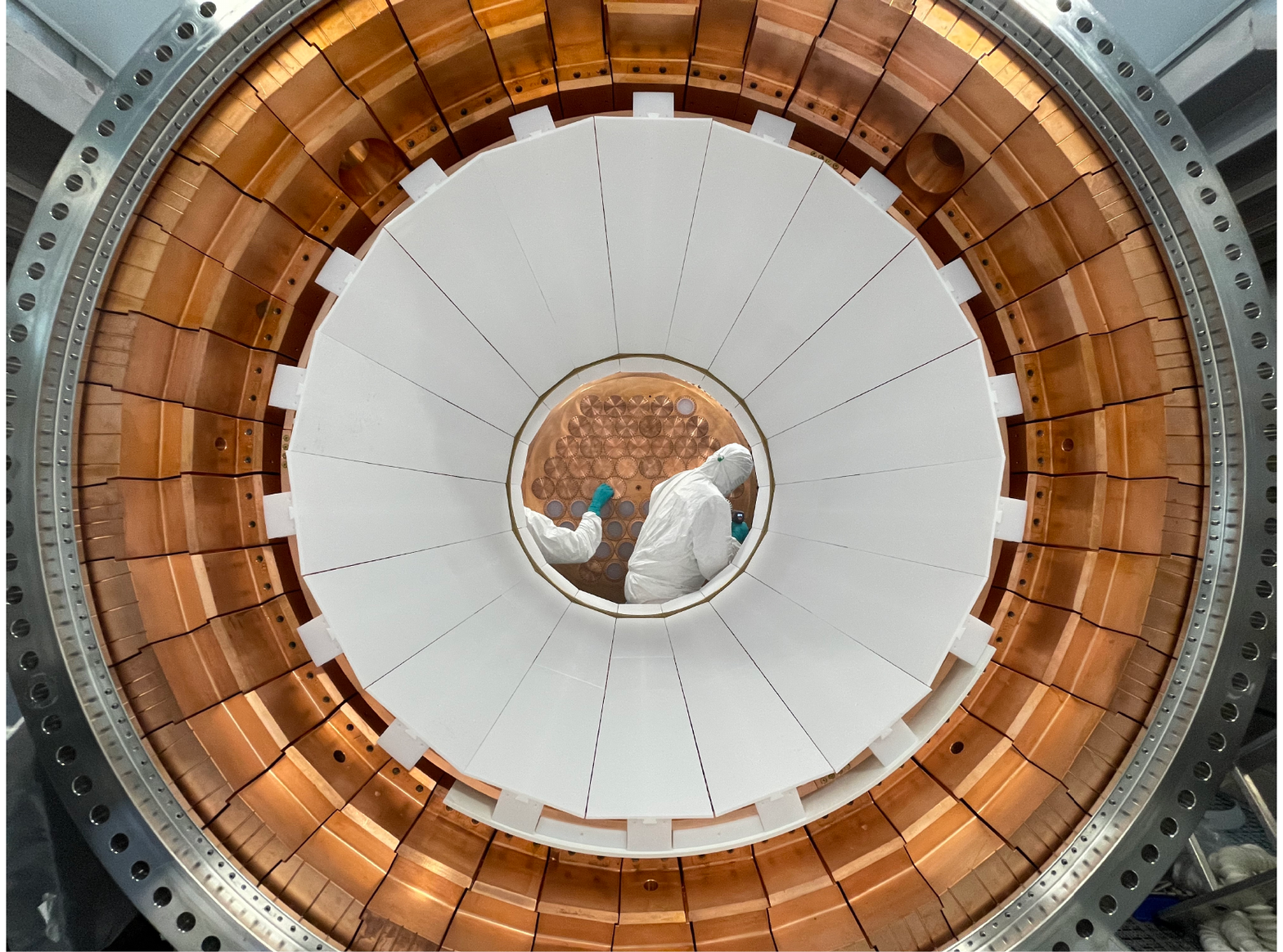}}
        \caption{}
        \label{fig:FC-wteflon}
    \end{subfigure}
  \caption{(a) Illustration of the NEXT-100 setup in Hall A of the LSC, showing the high-pressure time projection chamber, pressure vessel, internal components, and external shielding. (b) Operation principle of the high pressure xenon detectors of the NEXT program: a time projection chamber, using electroluminescence as amplification. (c) A picture of the pressure vessel and inner copper shell taken from the energy plane side during assembly. (d) Field cage after being assembled and inserted over the inner copper shell inside the pressure vessel, picture taken from the energy plane side. (e) Field cage after inserting the PTFE panels that create the light tube, picture taken from the tracking plane side. }
  \label{fig:next100}
\end{figure*}

\section{Introduction}
\sloppy The observation of the neutrinoless double beta decay (\lessnuBB) would be a major discovery, implying the violation of lepton number conservation and revealing key aspects of neutrino properties, such as the origin of its mass and the matter-antimatter asymmetry in the universe \cite{0nubbReview}. The Neutrino Experiment with a Xenon TPC (NEXT) program is advancing high-pressure xenon gas Time Projection Chambers (TPC) with electroluminescent amplification {(HPXeTPC-EL)}. As detailed throughout this paper, this technology represents a promising candidate for large-mass detectors in the search for \lessnuBB~decays. The first phase of the program was focused on the construction, commissioning, and operation of two detectors called NEXT-DEMO and NEXT-DBDM \cite{DEMO1, DEMO2}. These two prototypes demonstrated the potential of the technology by showing excellent energy resolution (a corresponding 0.5$\%$ FWHM \cite{DEMO3} at the \lessnuBB~energy in $^{136}$Xe, $Q_{\beta\beta}$ = \qty{2457.8(0.4)}{\kilo\electronvolt} \cite{Redshaw:2007un}) and the topological capabilities to differentiate signal from background \cite{DEMO4}. NEXT-White operated during the second phase of the program (from 2016 to 2021) \cite{NEWDetector} at the Laboratorio Subterr\'aneo de Canfranc (LSC), validating the HPXeTPC-EL technology in a large-scale radiopure detector. It exhibited an energy resolution better than 1$\%$ FWHM at the  $Q_{\beta\beta}$ \cite{NEWEres1, NEWEres2}, contributed to a background model assessment \cite{NEWBkg1, NEWBkg2} (specifically identifying the radioactive budget of various components), and successfully differentiated between single and double electron tracks \cite{NEWTrack1, NEWTrack2}. This detector enabled the measurement of the two-neutrino double beta decay rate {(\nuBB)} \cite{NEW2nubb}, and set a limit on the half-life of \lessnuBB~of $T^{0\nu} > 1.3\cdot10^{24}\text{yr}$ at 90$\%$ C.L. \cite{NEW0nubb}.

The NEXT-100 detector (Fig.~\ref{fig:NEXT100_detector}) constitutes the third phase of the program. The detector began operations in May of 2024 at LSC, and it is currently taking data after a commissioning phase. It consists of a {HPXeTPC} using an electroluminescence (EL) region to amplify the signal, and it is designed to hold $\sim$\qty{70.5}{\kilo\gram} of $^{136}$Xe at \qty{13.5}{\bar} in the active volume. A HPXeTPC-EL is composed of three main sections, all aligned axially, as illustrated in Fig.~\ref{fig:HPXeTPC}. The first section is the \textit{Buffer Region}. This region, positioned between a cathode at negative high voltage (up to \qty{-70}{\kilo\volt}) and an electrical ground shielding, prevents sparking and unwanted electroluminescence. The second section, the \textit{Drift Region} extends between the cathode and gate, maintaining a uniform electric field (up to $\sim$\qty{400}{\volt\per\centi\meter}). The final section is the \textit{Electroluminescence Region} of the detector. An intense electric field created in a very small (around \qty{10}{\milli\meter}) region between the TPC gate held at high voltage (up to \qty{-20}{\kilo\volt}) and the grounded anode, accelerates the ionization electrons sufficiently to produce a scintillation light. The EL field (from \qtyrange{1}{5}{\kilo\volt/\centi\meter/\bar}) is sufficient to induce EL but low enough to avoid additional gas ionization, which would affect the linearity of the amplification process \cite{ELAmpl}. The whole detector barrel is encapsulated with internal and external shielding in a high pressure vessel (Fig.~\ref{fig:VesselICS}). \\
The principle of operation of a HPXeTPC-EL is depicted in Fig.~\ref{fig:HPXeTPC}. When a charged particle interacts in high pressure gas it loses energy by ionization and by exciting the atoms in the gas. After de-excitation, these atoms emit vacuum ultraviolet (VUV) photons of \qty{172}{\nano\meter} wavelength creating a \textit{primary scintillation signal} (called S1, depicted with blue lines on Fig.~\ref{fig:HPXeTPC}). The S1 signal provides the information of the start time of the event ($t_{0}$). The secondary electrons created by ionization drift towards the anode under the influence of the electric drift field. The ionization signal is amplified by the EL and produces \textit{secondary scintillation light} (called S2, represented with red lines on the same figure). The light signal of S1 and S2 propagates along a reflective light tube and is collected through two different sensor planes on each end of the barrel. One plane is instrumented with 53 photo-multiplier tubes (PMTs) placed behind the cathode region of the TPC, while the other is instrumented with a plane of 3,584 silicon photo-multipliers (SiPMs) and collects the signal from the amplification region. The energy of the event is estimated by measuring the amplified S2 EL signal with the PMTs, the so-called Energy Plane (EP). This EP also records the previously mentioned prompt S1 signal. The plane of SiPMs (called Tracking Plane, TP) extracts the information from the drift electrons since the S2 light can provide the transverse ($x$, $y$) position. Thanks to the combination of both S1 and S2 information, the longitudinal ($z$) position of the event can also be obtained,
\begin{equation}
z = v_d \cdot (t_i-t_{0})    
\end{equation}
where the origin ($z=0$) is defined to be the edge of the EL region, $t_i$ is the time of each of the electrons making it to the EL region (S2 time), and $v_d$ the drift velocity corresponding to field conditions. The design of the NEXT detectors is able to retrieve the energy and position ($x$,$y$,$z$) information of the particles interacting on the detector with high accuracy. \\

The aim of the NEXT-100 detector is to reach a sensitivity to the \lessnuBB~half-life in the order of $\sim$\qty{e25}{\text{yr}} at 90$\%$ C.L., for a background index below \qty{1e-3}{\text{counts}\per(\kilo\electronvolt\cdot\kilogram\cdot\text{year})} and an energy resolution of $\sim1\%$~FWHM at the $Q_{\beta\beta}$ of $^{136}$Xe. This will demonstrate the scalability of the HPXeTPC-EL technology and reinforce the case for a future ton-scale detector. This paper describes the design and performance of the NEXT-100 detector. The information regarding its pressure vessel and shielding is summarized in Section~\ref{sec:vessel}, while all the inner elements like the TPC and the sensor planes are described in Sections~\ref{sec:tpc}, \ref{sec:ep} and \ref{sec:tp}. Improvements of the data acquisition and the gas system with respect to the previous-generation detectors are covered in Sections~\ref{sec:daq} and \ref{sec:gas}, respectively. Additionally, details regarding the radioactive budget of the detector and performance and stability observed during its first year of operation are covered in Sections~\ref{sec:radiopurity} and \ref{sec:commissioning}. \\

\section{The detector vessel and shielding} \label{sec:vessel}

\begin{figure}[t!]
  \centering
    \begin{subfigure}{0.45\textwidth}
        \centering
        \resizebox{0.8\textwidth}{!}{%
  \includegraphics[width=1.0\linewidth]{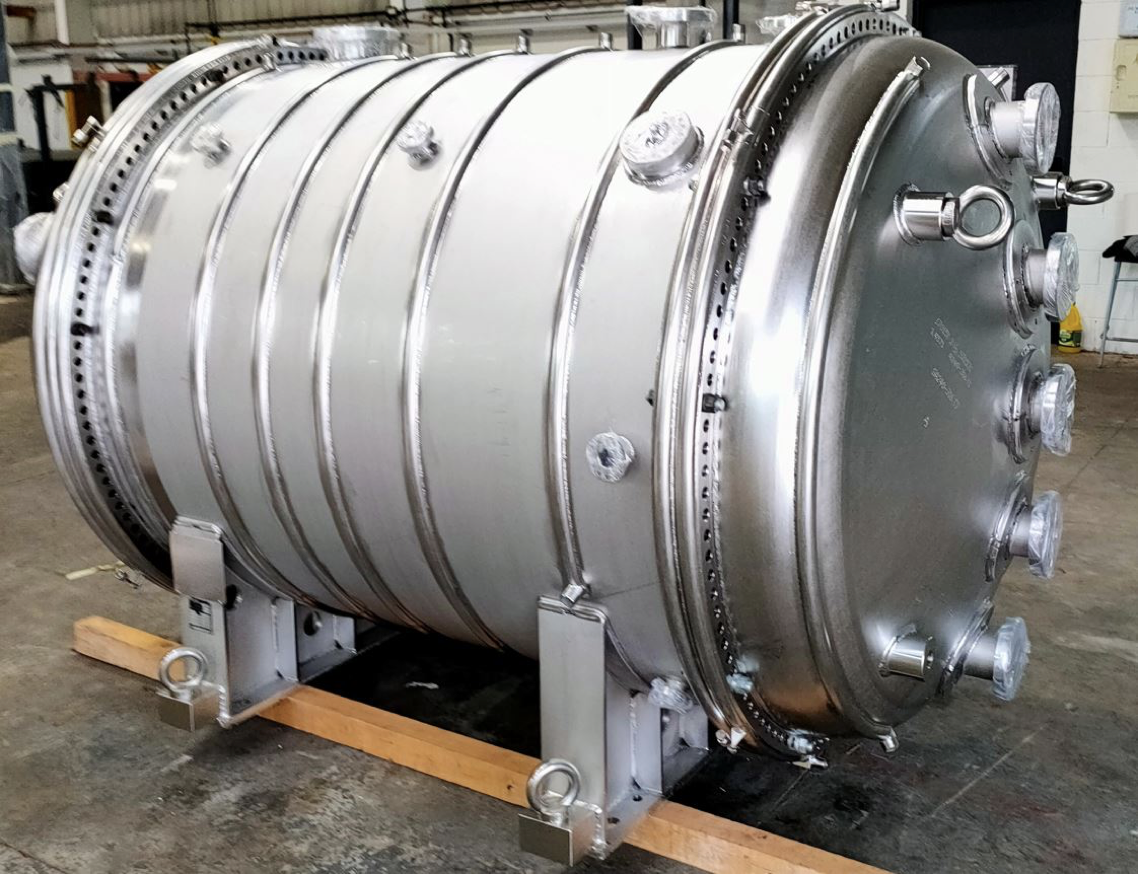}}
        \caption{}
        \label{fig:vessel}
    \end{subfigure}
    \begin{subfigure}{0.45\textwidth}
        \centering
        \resizebox{0.8\textwidth}{!}{%
  \includegraphics[width=1.0\linewidth]{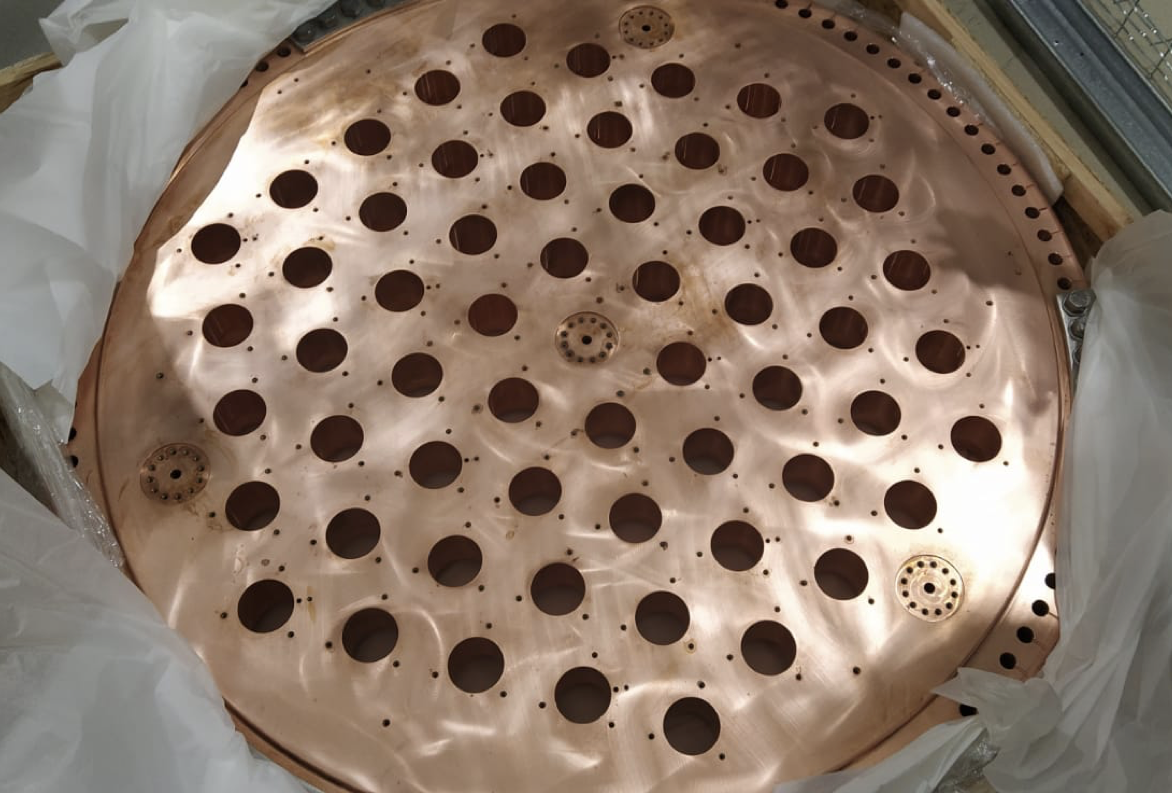}}
        \caption{}
        \label{fig:EP_copper}
    \end{subfigure}
    \begin{subfigure}{0.45\textwidth}
        \centering
        \resizebox{0.8\textwidth}{!}{%
  \includegraphics[width=0.8\linewidth]{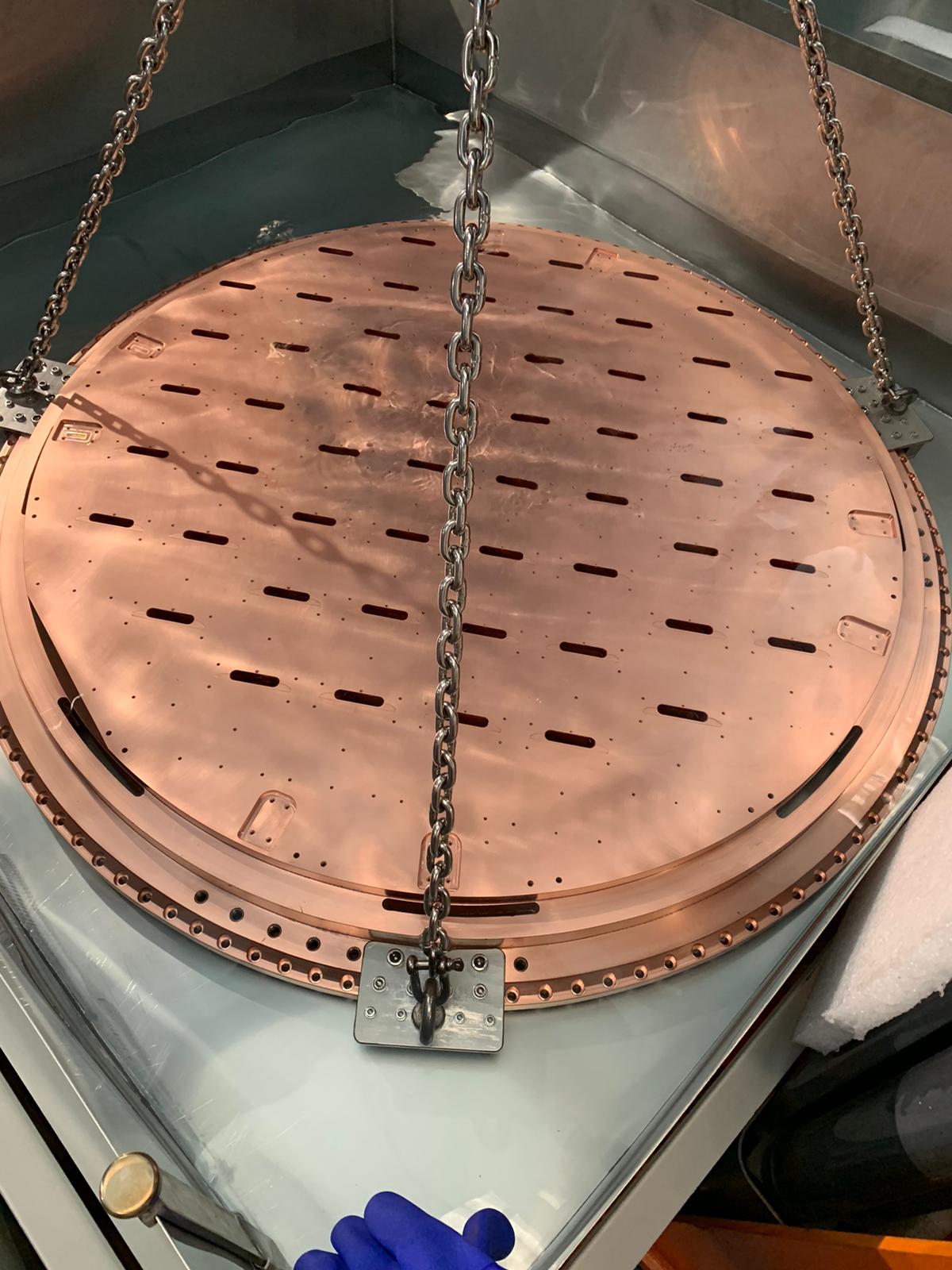}}
        \caption{}
        \label{fig:TP_copper}
    \end{subfigure}
  \caption{(a) Pressure vessel after polishing and prior to installation at the LSC. Inner Copper Shielding of (b) the Energy Plane and (c) the Tracking Plane. These pictures were taken after their machining and during the cleaning process, prior to installation in the NEXT-100 detector.}
  \label{fig:ICS}
\end{figure}

The NEXT-100 detector operates in the Hall A of LSC, inside a pressure vessel (Fig.~\ref{fig:vessel}) with a volume of \qty{1.7}{\meter}$^3$ which is fabricated with a stainless steel radiopure titanium alloy (316Ti) and can hold a maximum pressure of \qty{15}{\bar}. Several ports and connections are distributed along this vessel for mechanical and electronic integration of the gas system with the sensor planes. The xenon gas circulates through a dedicated gas system that keeps it continuously purified (details about the Gas System can be found in Section~\ref{sec:gas}). Several calibration ports for external radioactive sources are integrated into the detector. A total of four axial ports enable a full understanding of the detector performance thanks to their position along the drift region. Additionally, a calibration port for the use of an internal Krypton ($^{83\text{m}}$Kr) source is placed within the gas system (described also in Section~\ref{sec:gas}). The whole setup is placed on the same seismic table as the NEXT-White \cite{NEWDetector} detector (Fig.~\ref{fig:NEXT100_detector}), located on top of an elevated grating platform. \\
The NEXT-100 detector is shielded from external backgrounds through two layers. An outer shielding made of lead bricks (\textit{outer lead shielding}, depicted in Fig.~\ref{fig:NEXT100_detector}) reduces background gamma radiation from cosmic ray interactions and natural radioactivity from the LSC. The same outer lead shielding was used during the operation of NEXT-White detector. However, its characteristic red paint had to be removed since it had been observed during the operation of NEXT-White that it had a high radioactive component. An inner shielding, made with an ultra-pure copper shielding (ICS) of \qty{120}{\milli\meter} thickness, is designed to block lower-energy X-rays and secondary radiation. The barrel region of the ICS was designed with 40 individual copper blocks (Fig.~\ref{fig:VesselICS}), each one covering between 6-12 degrees of the total circumference to facilitate machining and assembly. The EP and the TP copper plates (\qty{140}{\milli\meter} and \qty{120}{\milli\meter} thick respectively) were machined as a single piece with the corresponding insertions required for sensors: holes for the coupling of PMTs coupling to the gas (Fig.~\ref{fig:EP_copper}) and elongated slots for SiPM cables (Fig.~\ref{fig:TP_copper}), respectively. The ICS of both caps have the same shielding purpose for the detector, but are coupled to the detector differently. The TP ICS holds the SiPMs and the EL region, and does not provide any isolation to the pressure volume. However, the EP ICS acts as a separation flange between the pressure volume and a vacuum region. This maintains the right operation conditions of the PMTs: vacuum and a dark environment.

\begin{figure}
    \centering
    \begin{subfigure}{0.45\textwidth}
        \hspace{5mm}\rotatebox{0}{\resizebox{0.90\textwidth}{!}{%
  \includegraphics[width=1.0\linewidth]{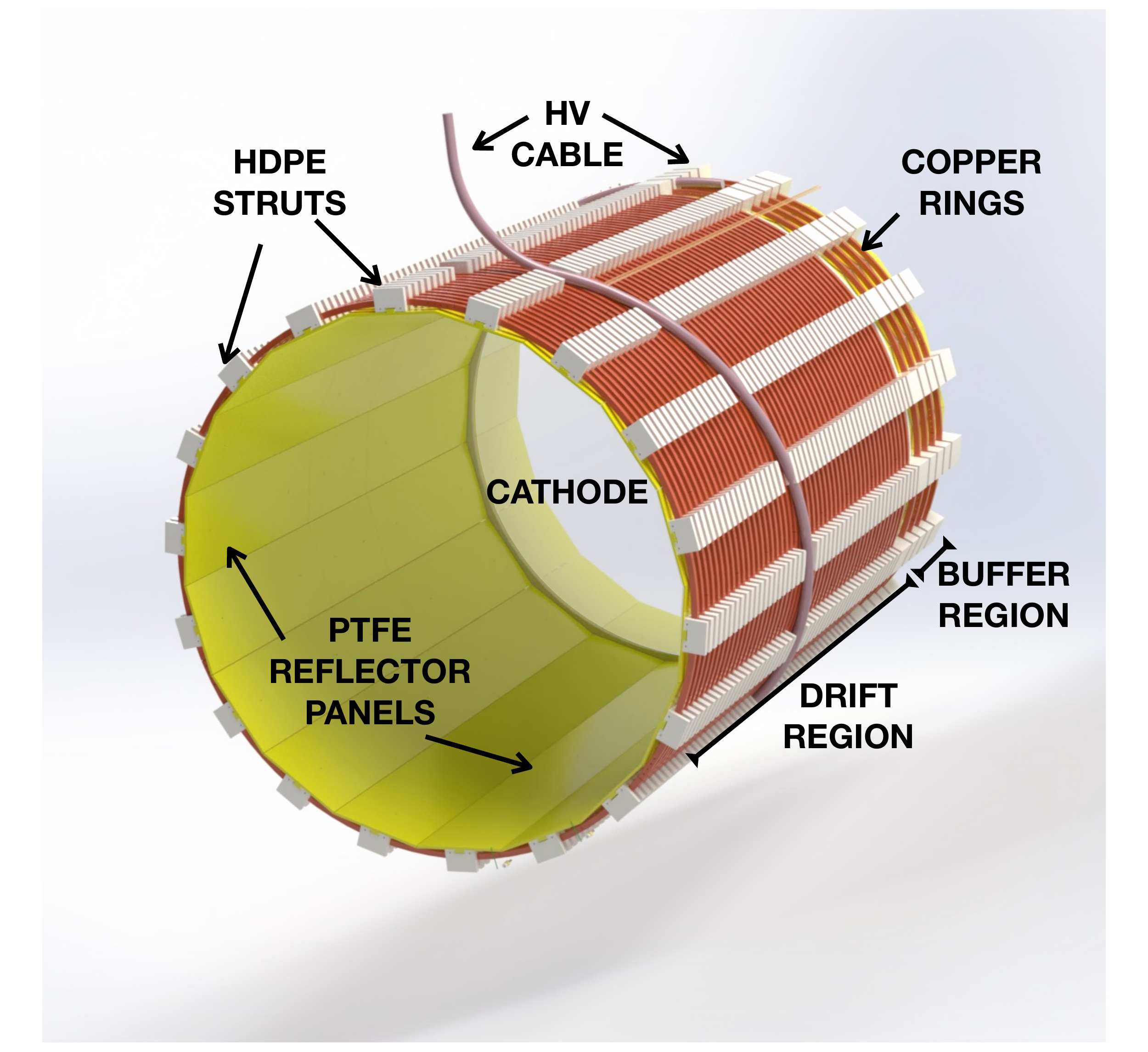}}}
        \caption{}
        \label{fig:FC-long}
    \end{subfigure}
    \begin{subfigure}{0.45\textwidth}
        \centering
        \resizebox{1.0\textwidth}{!}{%
  \includegraphics[width=1.\linewidth]{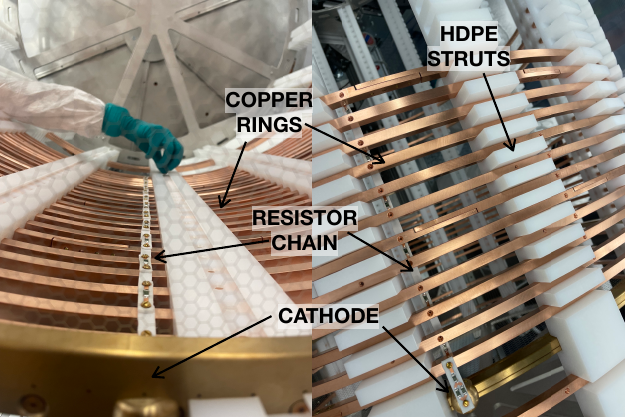}}
        \caption{}
        \label{fig:FC_resistors}
    \end{subfigure}
    \caption{(a) Drawing of the field cage without the outside insulator panel. The copper rings (red-brown) are inserted into the struts, the cable (pink) wraps around from the field cage connecting to the cathode (white), and the reflector panels (yellow) are slid into place through the struts. (b) Field cage during assembly. Resistor chains are visible along the drift region and connected to the cathode ring (left). Stainless steel wheels were used to help during assembly and insertion, and removed once it was inserted. }
    \label{fig:FC}
\end{figure}

\section{Time Projection Chamber}
\label{sec:tpc}

The NEXT-100 Time Projection Chamber (TPC) consists structurally of the field cage (FC), the cathode and EL region formed by a set of grids, and the high voltage feed-throughs (HVFT). This section details the design of the TPC, its components, and the assembly process. All materials were carefully selected to meet radiopurity requirements (more details in Section~\ref{sec:radiopurity}), with extensive ICP-MS and HPGe assays to limit contamination from $^{238}$U and $^{232}$Th.

\subsection{Field Cage}

The NEXT-100 Field Cage (Fig.~\ref{fig:FC-long}) main structure is held together by 18 high-density polyethylene (HDPE) struts that run along the full length of the FC. These plastic struts provide structural support (in Fig.~\ref{fig:FC-noteflon} and \ref{fig:FC-long}), space the copper field shaping rings, and are also used for fastening the light reflecting panels (Fig.~\ref{fig:FC-wteflon}) and cathode into place. The 52 field shaping copper rings serve as the circumferential mechanical support holding the weight of the whole FC. The drift region is formed by 48 rings that are spaced axially with a pitch of \qty{24}{\milli\meter}. They are built in three separate pieces and connected together with brass bolts to form a full ring. Each copper ring is attached to the grooves of the struts by nylon bolts. 
The cathode is supplied with high voltage (HV) via a cable connected at its upper section (more details about the connection are described in Sec.~\ref{Sec:HVFT}).
The HV cable makes a helical path path around the FC (Fig.~\ref{fig:FC-long}) from the cathode connection towards the gate region, where the HV feed-through is located. Ridges are cut into the HDPE to provide room and a path for the cable to lay. \\
The potential along the drift region is shaped by 3 resistor chains that are placed isometrically (visible in 3 positions in Fig.~\ref{fig:FC-noteflon}: top-right, bottom, and left-middle areas), connecting each copper ring and are attached to the cathode. Resistors from Vishay Techno with \qty{100}{\mega\ohm} resistance (CHRV100MEDKR-ND) were selected to ensure low currents in the FC and fulfill the radiopurity criteria. Circuit boards are made of CuFlon, a high-performance dielectric material made of polytetrafluoroethylene (PTFE) bonded to a copper foil. The PTFE used is unfilled, meaning it has no glass or ceramic fillers, giving it superior electrical properties and capable of meeting the radiopurity budget. Each resistor board, of dimensions \qtyproduct{28 x 6 x 2.7}{\milli\meter} connects two copper rings (Fig.~\ref{fig:FC_resistors}). This chain of resistors is mounted on the inside of the FC rings to avoid any possible sharp point from the resistors or the soldering in the high field region between the rings and the ICS. 
The final ring is connected to the cathode with a resistor board of \qty{150}{\mega\ohm} and a silicon bronze button similar to the one used for the HV cable connection (Section~\ref{Sec:HVFT}). Resistors were tested prior to installation, and after assembly connection to the rings was measured using a picoammeter. The FC was assembled outside the pressure vessel, on top of a cradle structure tooled to provide support during assembly and to push the main structure into the main pressure vessel.

\begin{figure}[t!]
  \centering
    \includegraphics[width=1.0\linewidth]{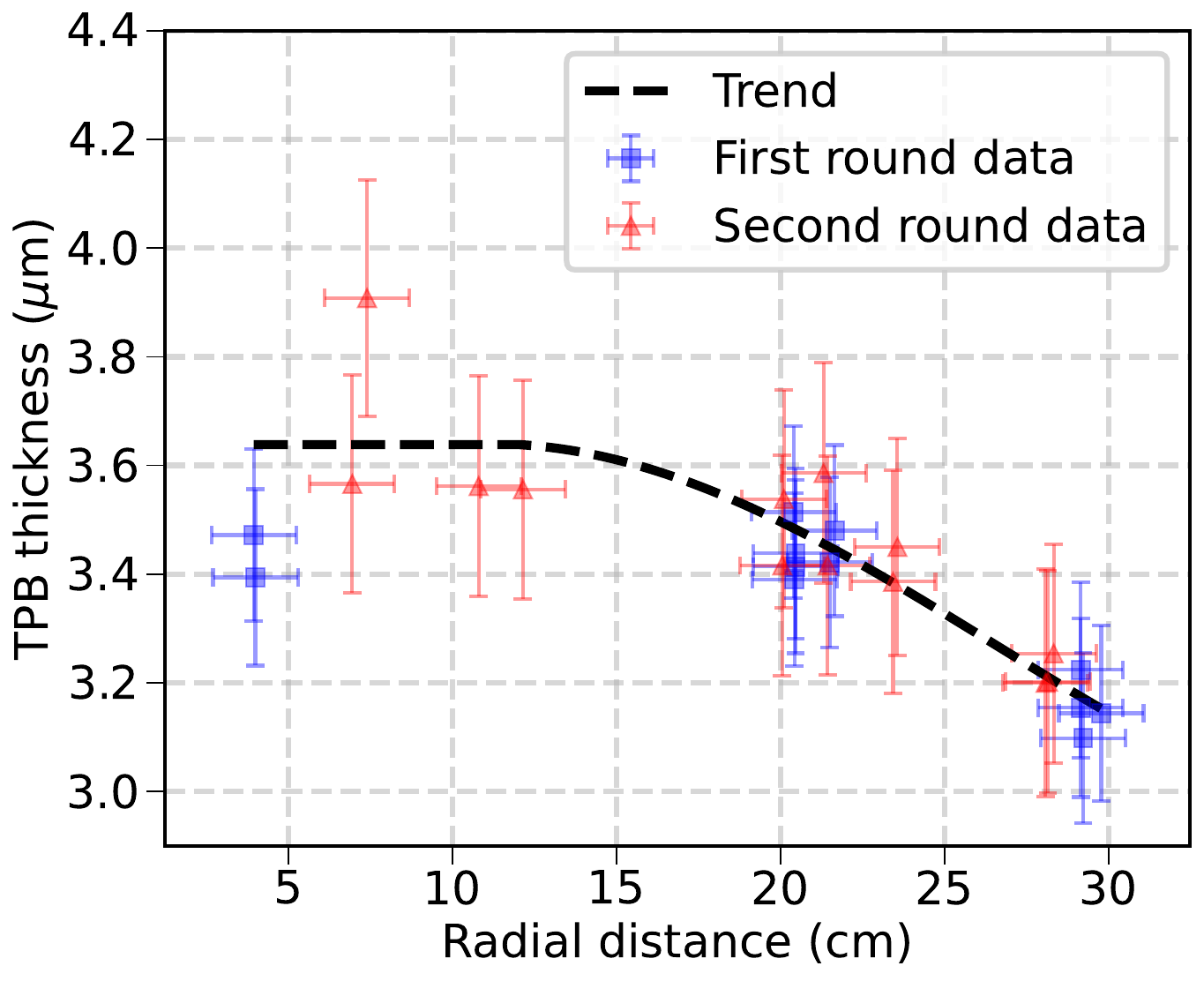}
    \caption{Expected thickness of the TPB coating of the PTFE panels covering the interior side of the light tube. Two test evaporations were evaluated (blue and red markers), showing an agreement at different radial distances within the evaporator (shown by the general trend in black). Values show a homogeneous coating between \qtyrange{3}{4}{\micro\meter} of TPB, ensuring a maximal efficiency for light emission  \cite{TPBStudies}. Thickness was estimated by positioning test slides where the panels would be placed and, after a regular evaporation, measuring the deposition over said slides with a stylus profilometer. Errors reflect the actual size of the slides (for the radial distance) and the variation of the coating along each slide (for the thickness).}
    \label{fig:coating}
\end{figure}

Panels of PTFE reflectors are attached to the struts with a dovetail connection and are coated with tetraphenyl butadiene (TPB) (Fig.~\ref{fig:FC-wteflon} and \ref{fig:FC-long}), which is a wavelength shifter. These reflectors create a light-tube that delimits the active volume and enhances photon collection efficiency. The PTFE panels have a thickness of \qty{5}{\milli\meter} in their flat region, optimizing maximum light reflectivity with the minimum amount of material \cite{PTFEStudies}. The NEXT-100 light tube has been designed in panel sections along the longitudinal axis to simplify their coating and assembly. Smaller plastic pieces allow for easier TPB coating and transportation, and reduces the stress on the detector during vacuum and pressure cycles from expansion of the PTFE. Additionally, this arrangement was meant to test the design and construction of larger future detectors. Prior to coating, the panels were sanded in order to ensure a homogeneous surface and to remove specular reflections \cite{PTFEStudies}, then cleaned to remove any surface contamination. The quality of the TPB deposition was monitored and measured during evaporations using profiles at different positions along the evaporator. A homogeneous layer of \qtyrange{3}{4}{\micro\meter} (Fig.~\ref{fig:coating}) along the whole FC surface was observed, ensuring an optimum transformation and reflection of the xenon light \cite{TPBStudies}. Panels were transported to LSC in dark sealed bags filled with a nitrogen atmosphere and were the last piece installed in the detector to minimize the degradation of the TPB when exposed to light. The dimensions of each of the components of the field cage are described on the Table~\ref{table:fc_dimensions}.\\

\begin{table}[t!]
\centering
\caption{Dimensions of the individual components of the field cage, where the light-tube drift region defines the active volume of the detector.}
\label{table:fc_dimensions}
\begin{tabular}{c|c|c}
\hline
 \textbf{Component} & \textbf{Length [mm]}  & \textbf{Diameter [mm]} \\ 
\hline
HDPE Insulator & 1427 & 1105  \\
HDPE Struts & 1427 & --  \\
Copper Rings & -- & 1014  \\
Light Tube Buffer Region & 241 & 983 \\
Light Tube Drift Region & 1187 & 983  \\
\hline 
\end{tabular}
\end{table}

An outer layer made of two pieces of HDPE serves as an insulator separating the FC from the internal copper shielding within the vessel. They are mechanically designed as a flat sheet of quarter-inch thickness, wrapped directly around the external FC structure to prevent sparking between the field-shaping copper rings and the ICS. The inner layer, directly touching the FC, covers the whole surface. The outer layer in contact with the copper shielding is designed to protect the gap from the inner layer and avoid any direct contact with the copper rings. This two layer outer insulator was slid over the copper layer before the FC was inserted into the pressure vessel. 

\subsection{Cathode and Electroluminescence Meshes}

The NEXT-100 detector drift field and amplification region are produced by photoetched tensioned hexagonal stainless-steel meshes of \qty{1}{\meter} of diameter. Full details of their design, characterization, and installation can be found in Ref.~\cite{N100Mesh}. These have been tested to ensure uniformity and stability for the high-precision energy measurement in the detector. 

\begin{figure}[t!]
  \centering
  \includegraphics[width=0.9\linewidth, angle=0]{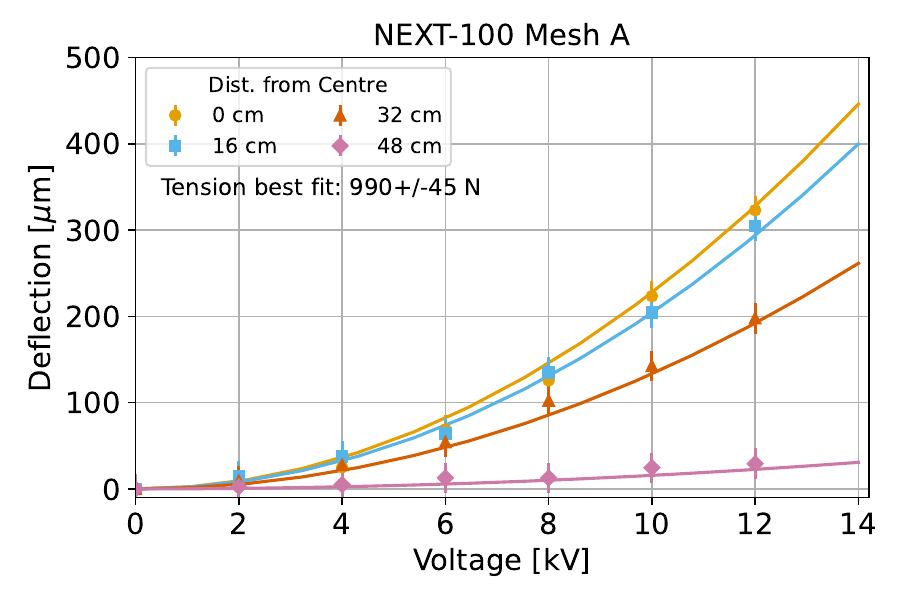}
  \caption{Measurement of the electrostatic deflection as a function of applied voltage shown for one of the EL meshes. Measurements are fit to a deflection model and the extracted mesh tensions are 990$\pm$45~N and 835$\pm$40~N \cite{N100Mesh}.}
  \label{fig:DeflectionMesh}
\end{figure}

\begin{figure*}[t!]
    \centering
    \begin{subfigure}{0.45\textwidth}
        \centering
        \includegraphics[width=1.0\linewidth]{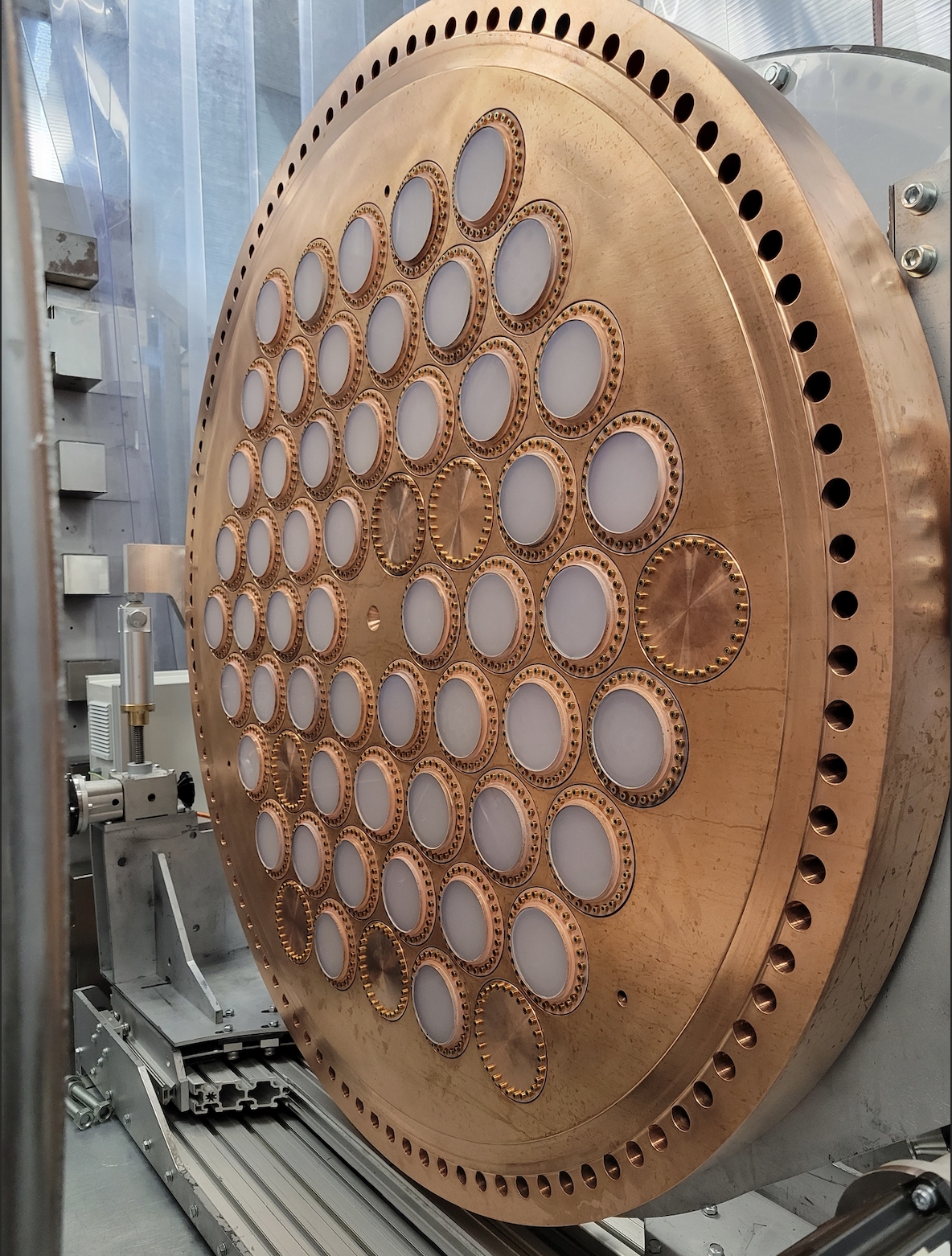}
        \caption{}
        \label{fig:EP}
    \end{subfigure}
    \begin{subfigure}{0.45\textwidth}
        \centering
        \includegraphics[width=0.98\linewidth]{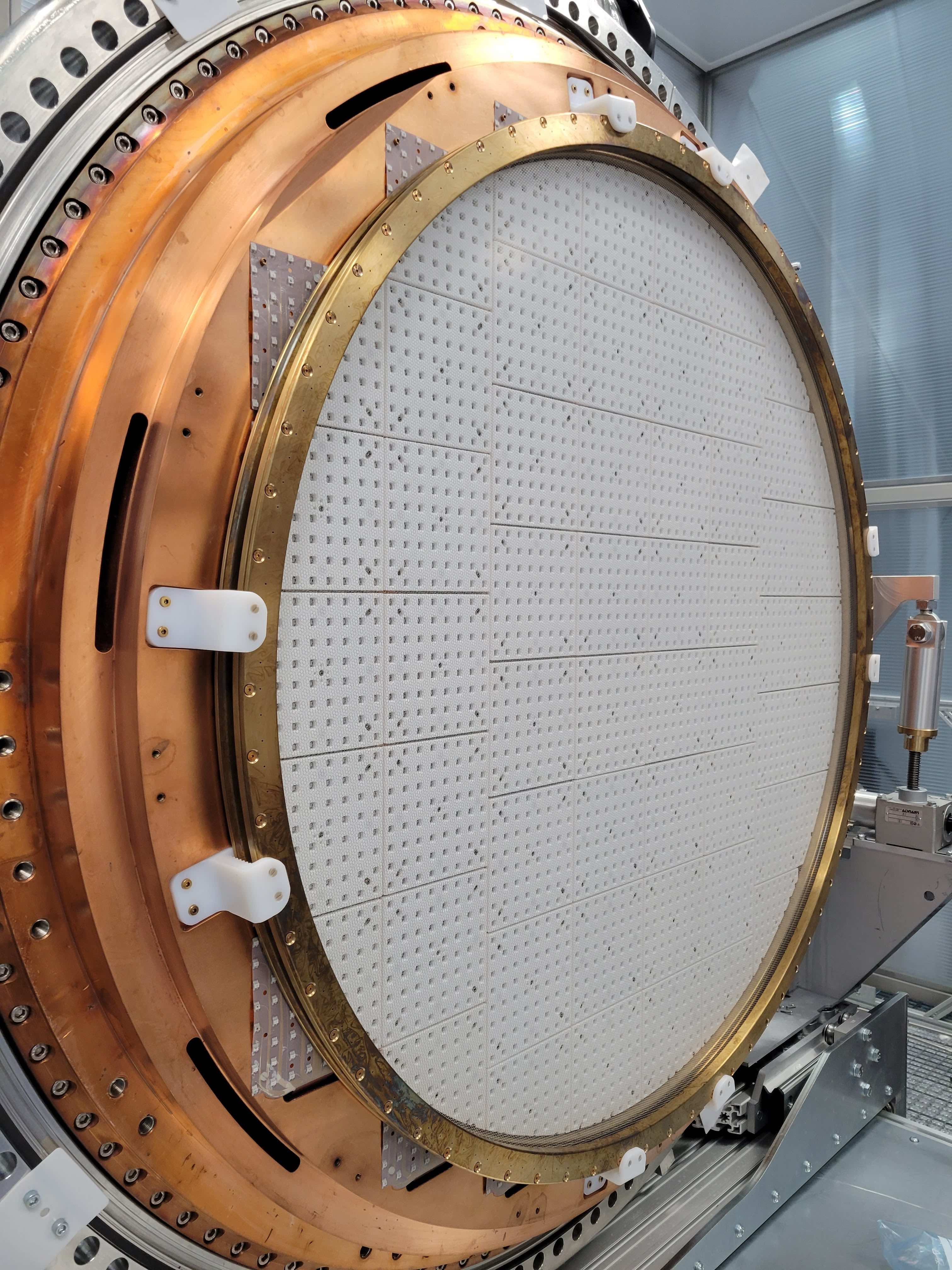}
        \caption{}
        \label{fig:TP}
    \end{subfigure}\\
    \begin{subfigure}{0.20\textwidth}
        \centering
        \includegraphics[width=0.8\linewidth]{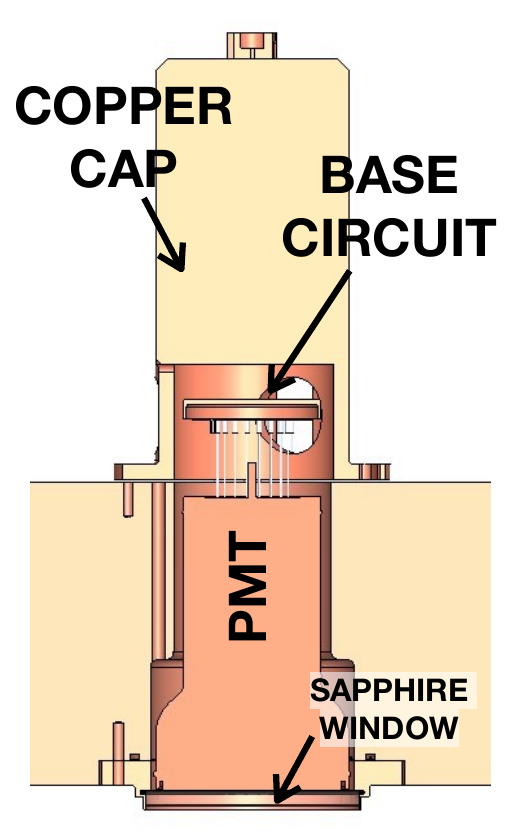}
        \caption{}
        \label{fig:EP_LED_hole}
    \end{subfigure}
    \begin{subfigure}{0.25\textwidth}
        \centering
        \includegraphics[width=1.0\linewidth]{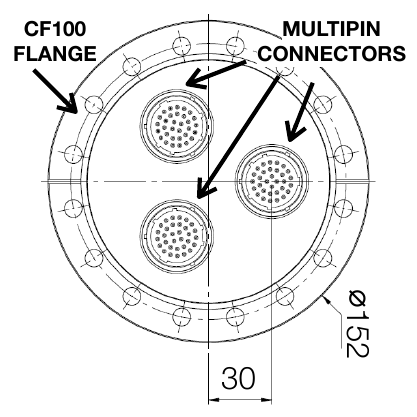}
        \caption{}
        \label{fig:EP_FT}
    \end{subfigure}
    \begin{subfigure}{0.23\textwidth}
        \centering
        \includegraphics[width=0.95\linewidth]{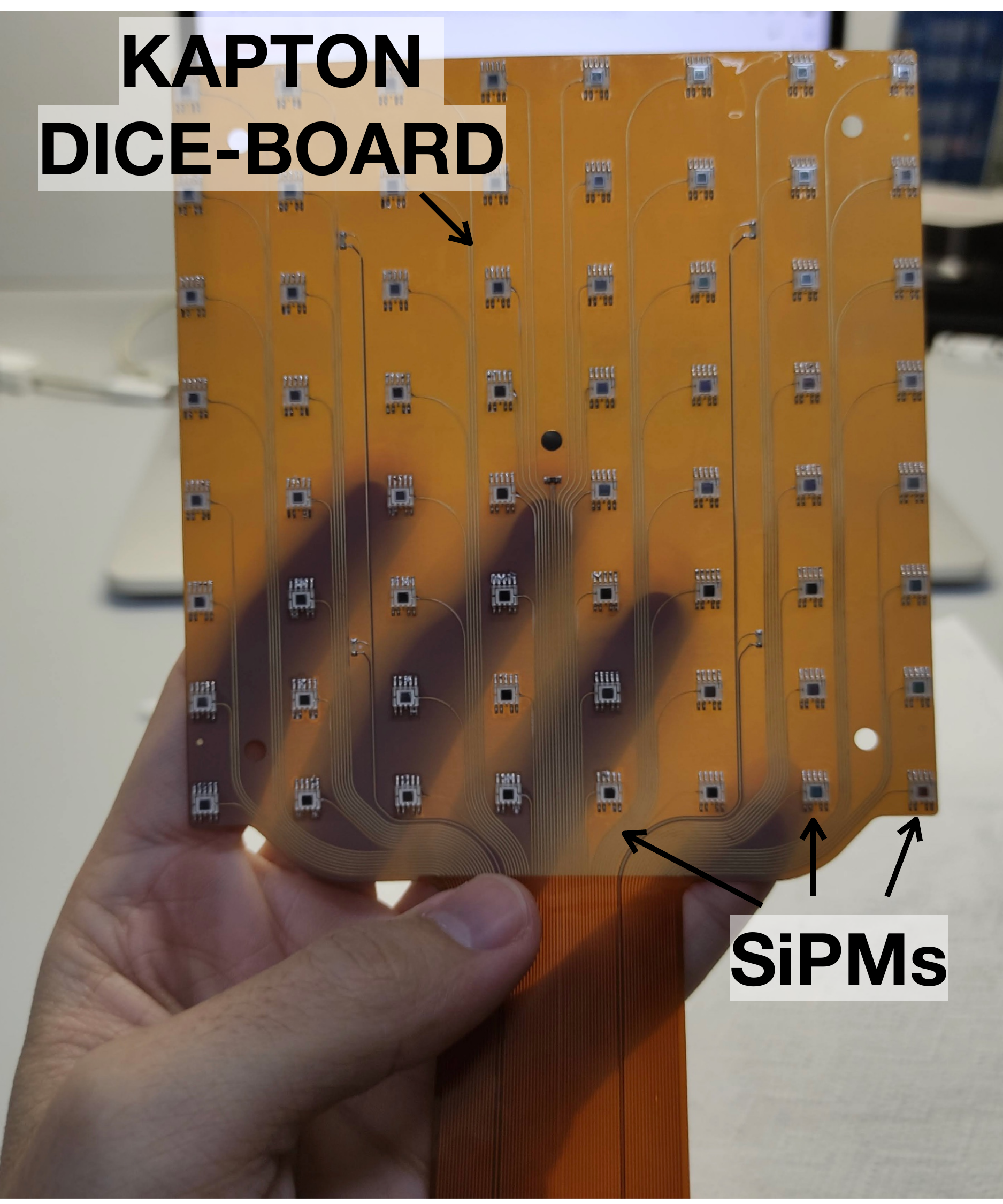}
        \caption{}
        \label{fig:TP_kapton}
    \end{subfigure}
    \begin{subfigure}{0.23\textwidth}
        \centering
        \includegraphics[width=0.85\linewidth]{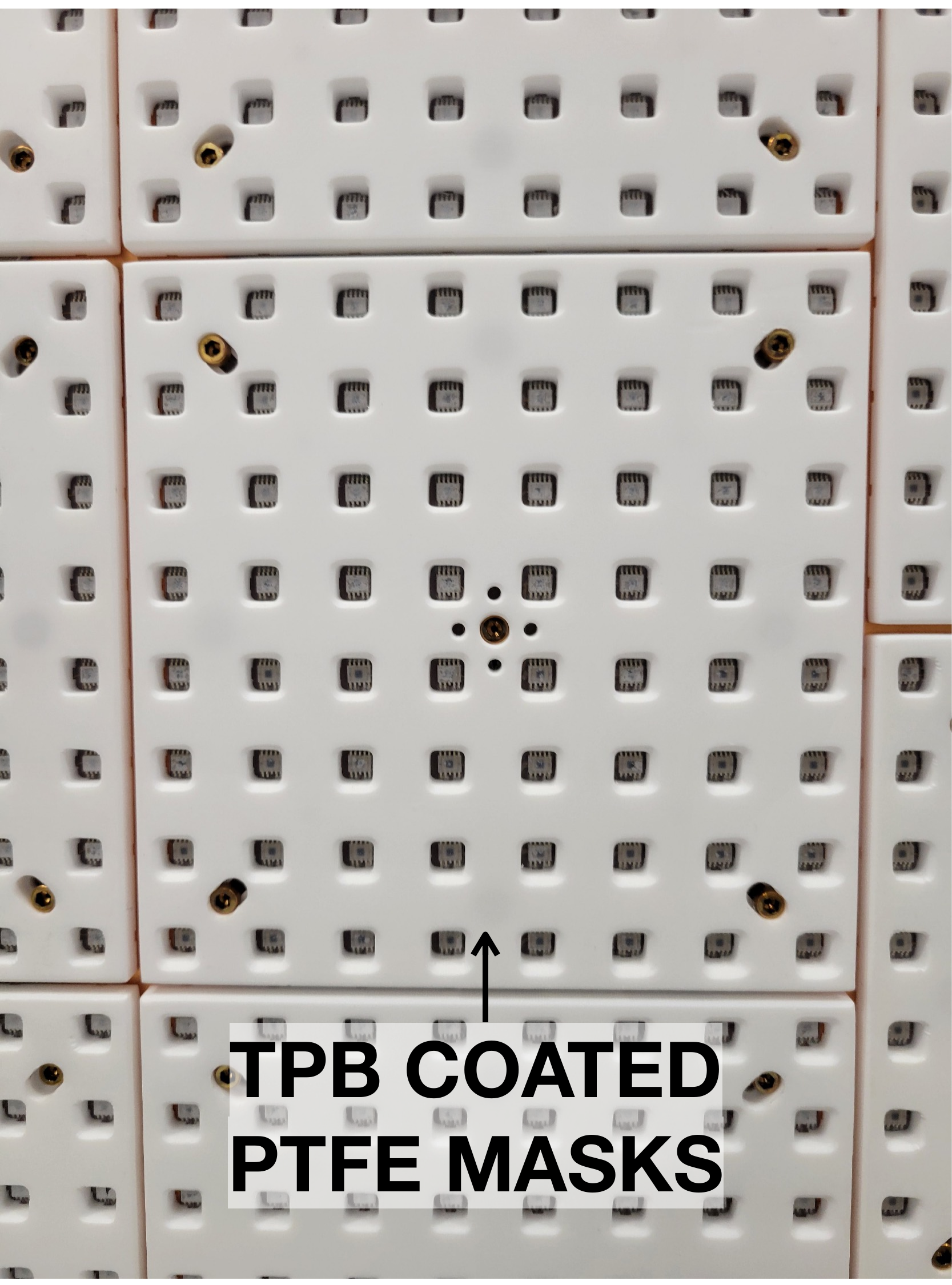}
        \caption{}
        \label{fig:TP_TPB}
    \end{subfigure}
    \caption{(a) Energy plane of the NEXT-100 detector during assembly. The PMTs are coupled to the xenon pressure region through sapphire windows. Copper covers were used during the pressure tests, and replaced by sapphire windows and coupled PMTs during assembly. (b) Tracking plane of the NEXT-100 detector after assembly. 56 DICE-boards assembled on the copper plate with PTFE masks on top. The EL gap is also assembled on top of the TP copper plate using HDPE brackets. (c) 3D drawing of the PMT cap and the LED hole at the copper plate. (d) Drawing of the custom feedtroughs used for the PMTs. (e) Kapton DICE-Board with SiPMs wielded. The thin layer of Kapton circuit ensures enough rigidity for assembly and proper flexibility for the pigtail cables. (f) DICE-Board assembled on the tracking plane, with PTFE mask on top of it.}
\end{figure*}

The EL region consists of two parallel meshes featuring a \qty{2.5}{\milli\meter} hexagonal pattern with a 90$\%$ optical transparency. This is structurally placed over the TP copper plate using HDPE brackets that maintain the amplification gap (see Fig.~\ref{fig:TP}) of {(9.70$\pm$0.15)~mm} (measured after installation). Both meshes are held in place and tensioned with a silicon bronze frame that was chosen for its mechanical strength and radiopurity. Electrostatic deflection of the meshes under high voltage was measured (Fig.~\ref{fig:DeflectionMesh}) using an optical microscope system. The extracted tension was $~\sim$\qty{900}{\newton}, which results in deflections that remain below \qty{0.4}{\milli\meter} at the expected operating conditions. For pressures in the range {10-15~bar}, the expected impact of the electrostatic deflection on the total energy resolution does not vary more than 0.15\% FWHM for electrical fields greater than \qty{16}{\kilo\volt\per\centi\meter}. In addition, studies have shown that the EL mesh alignment (shifted, aligned, or rotated configurations) had minimal impact on the energy resolution \cite{N100Mesh}. Breakdown voltage tests in air and argon confirmed that the EL meshes withstand discharges from operating voltages up to \qty{28}{\kilo\volt} without structural damage \cite{N100Mesh}. Post-installation tests verified that the assembled system met the required mechanical and electrical specifications. 

\begin{figure}
    \centering
    \begin{subfigure}{0.45\textwidth}
        \centering
        \resizebox{1.0\textwidth}{!}{%
  \includegraphics[width=1.2\linewidth]{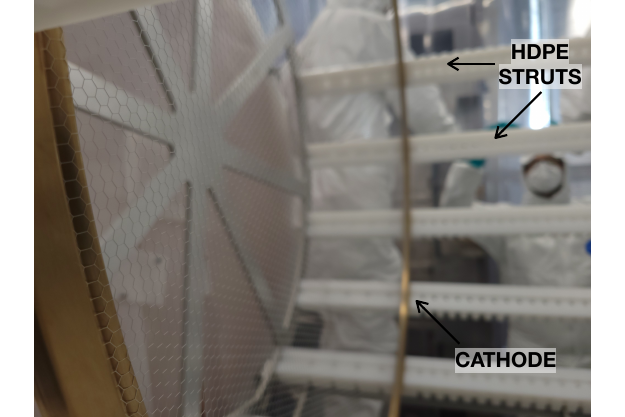}}
        \caption{}
        \label{fig:FC_cathode}
    \end{subfigure}
    \begin{subfigure}{0.45\textwidth}
        \centering
        \resizebox{0.9\textwidth}{!}{%
  \includegraphics[width=1.2\linewidth]{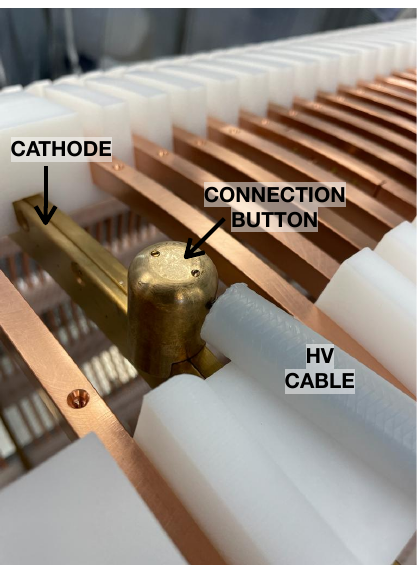}}
        \caption{}
        \label{fig:FC_cathodecable}
    \end{subfigure}
    \caption{(a) Cathode close up during the assembly of the field cage. (b) Connection of the high voltage cable with the cathode through a silicon bronze button. }
    \label{fig:Cathode}
\end{figure}

The cathode mesh features a \qty{5}{\milli\meter} hexagonal pattern to optimize transparency (95$\%$) and light collection efficiency, while minimizing electric field leakage. This mesh is designed to provide a constant drift field of $\sim$\qty{400}{\volt\per\centi\meter} in the active region, which implies a field of $\sim$\qty{2}{\kilo\volt\per\centi\meter} in the buffer region. The cathode operates at a lower local field strength compared to the $\sim$\qty{20}{\kilo\volt\per\centi\meter} of the EL, making the electrostatic deflection of the cathode mesh less significant. The cathode mesh is also tensioned with a silicon bronze frame. However, due to the lower fields applied less tension is required. The cathode is held on the main structure of the FC (see Fig.~\ref{fig:FC_cathode}). The conductive center of the high voltage cable connects to the cathode (see Fig.~\ref{fig:FC_cathodecable}) through a silicon bronze button that fits the mesh frame and is secured with brass screws. The material fulfills the same mechanical and radiopurity specifications of the mesh frame, and has a rounded shape to minimize electric fields arising from sharp points in the connection region.

\begin{figure}[t!]
  \centering
    \begin{subfigure}{0.45\textwidth}
        \centering
        \resizebox{1.0\textwidth}{!}{%
  \includegraphics[width=0.9\linewidth, angle=0]{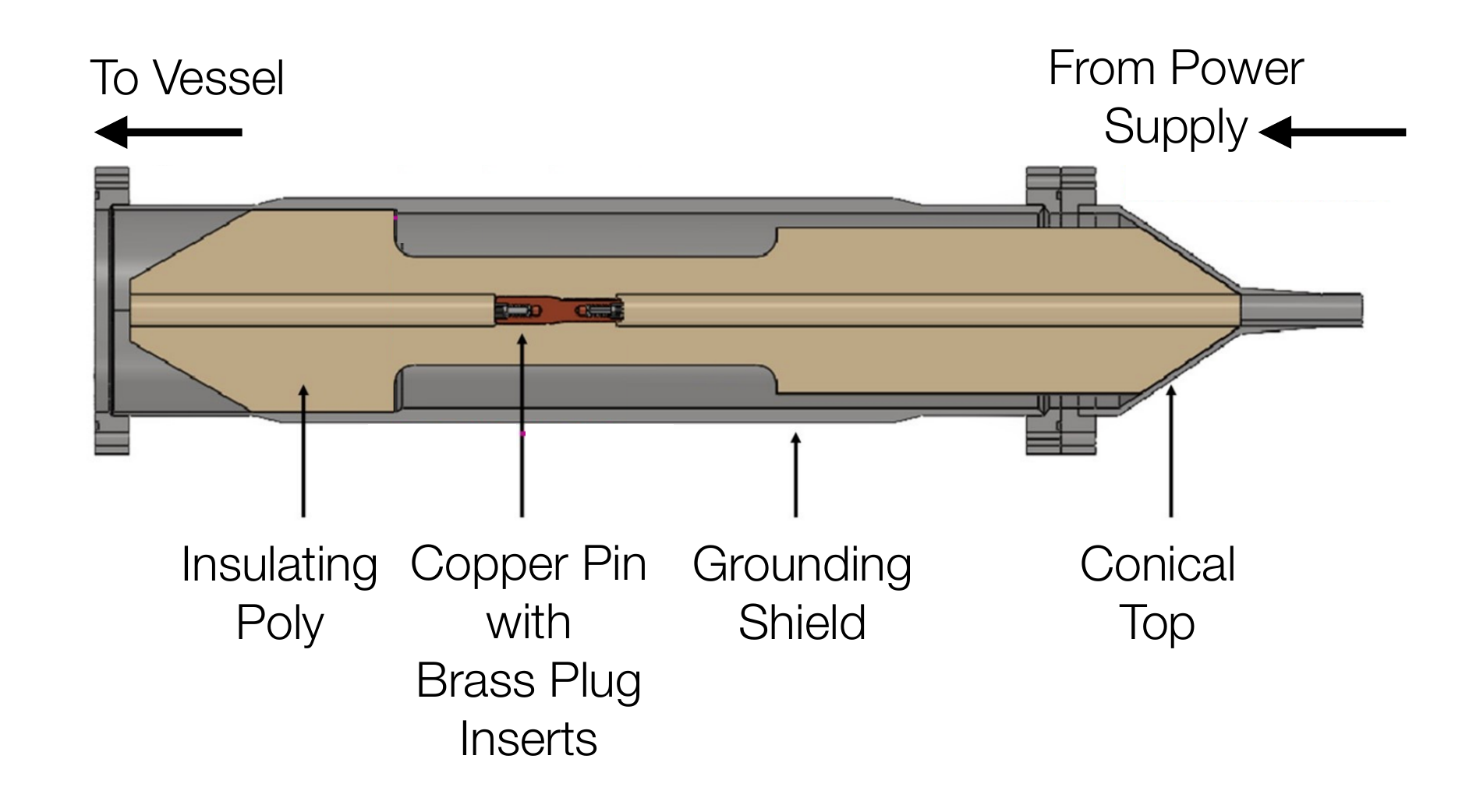}}
        \caption{}
        \label{fig:HVFT}
    \end{subfigure}\\
    \begin{subfigure}{0.45\textwidth}
        \centering
        \includegraphics[width=1.0\linewidth, angle=0]{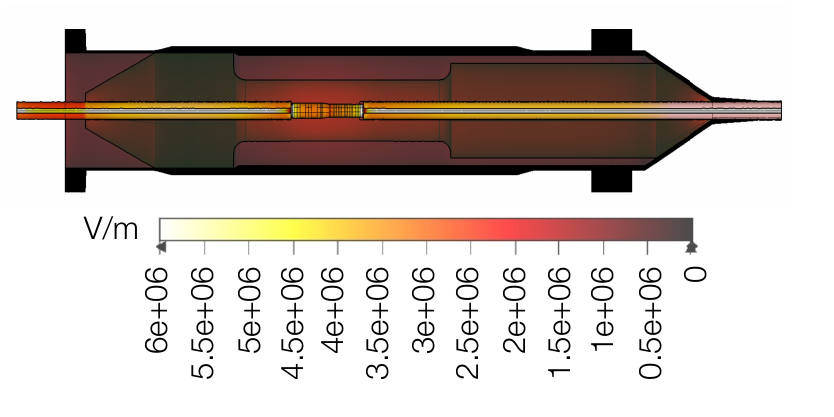}
        \caption{}
        \label{fig:HVFT_comsol}
    \end{subfigure}
  \caption{(a) Cross section of the feed-through for the high voltage connection of the cathode. More details can be found in \cite{HVFTPaper}. (b) Simulation of the electric fields using COMSOL within the feed-through when operating at 100 kV \cite{HVFTPaper}. }
  \label{fig:Voltage}
\end{figure}

\subsection{High Voltage Feed-throughs}\label{Sec:HVFT}
The high voltage of the NEXT-100 TPC is provided through three feed-throughs (FT) on the main vessel: a cathode HVFT, a FT connected to the last ring of the FC, and an EL gate FT. The three FTs are designed to deliver a stable voltage while maintaining insulation and preventing electrical breakdowns. Their design ensures durability based on tests of hydrostatic pressure (up to \qty{20}{\bar}) and vacuum tests (below \qty{1e-5}{\torr}). \\
The cathode HVFT (Fig.~\ref{fig:HVFT}) is designed to deliver a stable voltage up to \qty{-70}{\kilo\volt} to the cathode mesh. It is placed over the vessel on an upper port (see Fig.~\ref{fig:NEXT100_detector}) and provides connection to the cathode through the HV cable (Fig.~\ref{fig:FC-long}, \cite{HVFTPaper}). The outer grounding shield consists of a stainless steel 316 cylinder and it is connected to the vessel to provide a good ground connection. An electrical insulation sheath was designed with tapered ends to reduce electric field concentrations and minimize charge build-up (Fig.~\ref{fig:HVFT_comsol}). Internal and external cables are connected through an inner copper pin designed with interference fits to avoid air gaps and ensure stable electrical contact through a press-fit connection. The HV cables are attached using modified banana plug connectors. The grounding shield is extended using a metallic hose clamp over the cable’s shielding braid.\\

Since the EL meshes are not directly attached to the FC but are instead supported by the TP copper plate, it is essential to fully control the potential at the end of the drift region. To achieve this, a HV connection is established at the final ring of the FC through a FT. This FT consists of an outer stainless steel tube with a flange, connected through an inner PE1000 tube and a stainless-steel rod. This rod holds an inner copper tube with a single pin high voltage connector (GES; Type SB150 50 kVDC) to ensure a safe and stable voltage connection. The assembly of the FT was done with cryogenic shrink fitting, meaning that all the parts had to be machined with precise tolerances. After the machining and assembly, a leak and pressure test was done, which confirms that the FT can operate safely up to \qty{15}{\bar}. A FT with the same design is connected to the EL mesh that acts as a gate providing the desired voltage for amplification. Cathode voltage is supplied by a {FUG HCP 140-100000}, while the last ring and the gate are supplied by a {FUG HCP 35-35000}.\\

\begin{figure}[t!]
    \centering
    \begin{subfigure}{0.45\textwidth}
        \centering
        \resizebox{0.95\textwidth}{!}{%
  \includegraphics[width=0.8\linewidth, angle=0]{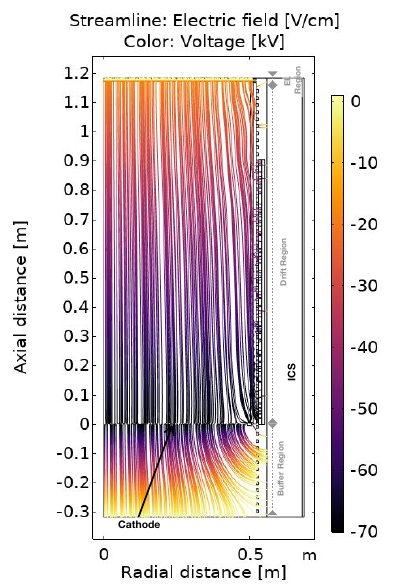}}
        \caption{}
        \label{fig:FC_Comsol}
    \end{subfigure}\\
    \begin{subfigure}{0.45\textwidth}
        \centering
        \resizebox{0.95\textwidth}{!}{%
  \includegraphics[width=1.0\linewidth]{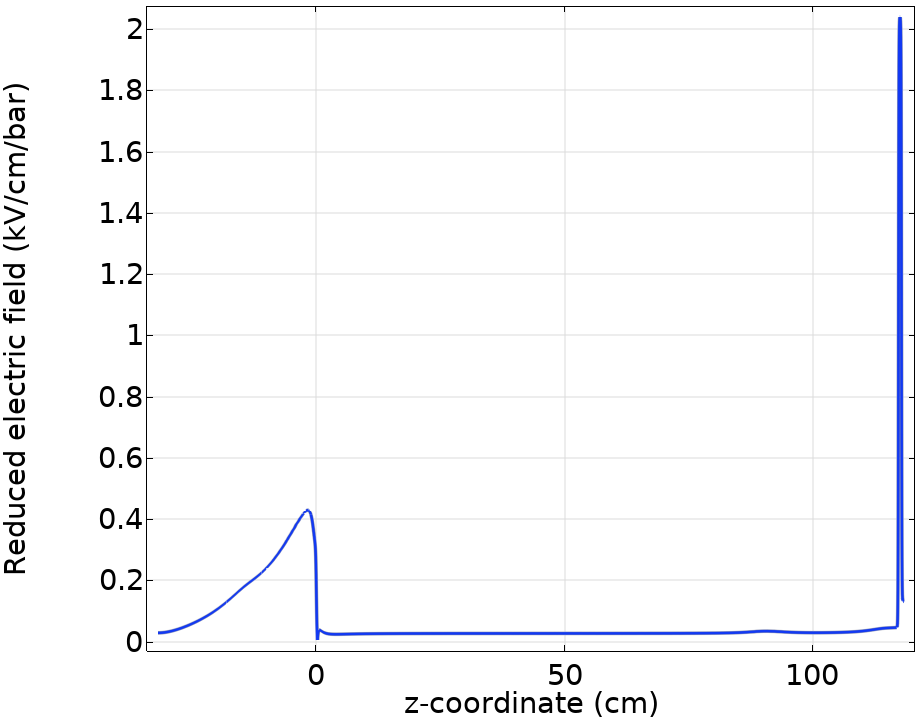}}
        \caption{}
        \label{fig:FC_ComsolWalls}
    \end{subfigure}
  \caption{(a) Simulations of the electric field in the NEXT-100 field cage using COMSOL at the maximum operation voltages: cathode at {-\qty{70}{\kilo\volt}} and gate at {-\qty{20}{\kilo\volt}}. Electric field lines are homogenous and constant along the detector drift region, corresponding to a electric field of $\sim$\qty{400}{\volt\per\centi\meter}. (b) Reduced electric field evaluated at \qty{20}{\milli\meter} from the PTFE walls, corresponding to the fiducial cut boundary where non-homogeneous effects are most significant.}
  \label{fig:Voltage}
\end{figure}

The electric field at different operation voltages has been simulated using COMSOL (Fig.~\ref{fig:FC_Comsol})  finite element analysis software, where a 2D-axis-symmetric approximation was used and radial components of the electric field have been evaluated through different axial positions. The field uniformity is satisfactory (within 10$\%$) for a radial distance of \qty{20}{\milli\meter} (Fig.~\ref{fig:FC_ComsolWalls}) from the teflon walls, showing no sharp points at different drift field values and ensuring the good quality of the electric field in the fiducial region.

\section{Energy Plane}\label{sec:ep}

The event energy (S2) and the primary scintillation light (S1) in the NEXT-100 detector is measured by the Energy Plane shown in Fig.~\ref{fig:EP}. It operates with 53 bialkali photocathode PMTs ({R11410-10} from Hamamatsu) with a surface diameter of \qty{64}{\milli\meter}, in a vacuum region with a gain of $5 \cdot 10^6$. The vacuum operated PMTs are coupled to the pressure volume through holes machined in the copper plate and closed with sapphire windows (Fig.~\ref{fig:EP_copper}). These windows are coated with poly ethylenedioxythiophene (PEDOT), a resistive and highly transparent compound that helps to define the ground at the EP while at the same time avoiding sharp electric field components and preventing charge accumulation on the windows. In addition, the sapphire windows are also coated with a thin layer of TPB to shift the direct VUV light from xenon to blue, where sapphire is highly transparent, maximising the total light collection. The PMTs are coupled to the sapphire windows with an optical gel (NOA76 from the company Edmund Optics). The design of the EP plate can accommodate up to 60 PMTs. However, some of the sapphire windows did not pass the quality tests needed to guarantee a safe operation of the detector. Simulations have demonstrated that a PMT coverage of approximately 25$\%$ is sufficient to achieve the excellent energy resolution required for the experiment, similar to that observed in the NEXT-White detector~\cite{NEWEres2}. Based on these results, the decision was made to operate the Energy Plane with only 53 PMTs.

Each PMT receives power through a base circuit built on a Kapton printed circuit board (PCB). As the PMTs are operating at high vacuum, the heat produced by the circuit needs to be dissipated by direct thermal contact. This was solved by covering the base with a copper cap (Fig.~\ref{fig:EP_LED_hole}) filled with radiopure epoxy that guarantees a good contact between the electronic components and the copper cap. The output of the PMTs is read in pseudo-differential mode by extracting the signal on the last and the third-to-last dynodes. This connection is made with twisted Kapton cables connecting the PMT bases with a feed-through in the torispherical head of the pressure vessel (visible in Fig.~\ref{fig:NEXT100_detector}). Electronic signals are retrieved from the pressure vessel through two custom feed-throughs each on CF100 flange. Each of them has 30 PMTs connected through 3 multipin connectors (Fig.~\ref{fig:EP_FT}). 

The Front-end electronics (FEE) is based on a differential transmission to effectively suppress noise coupled to the signal, which is essential for the detection of single photoelectrons and the identification of S1 signals. A differential amplifier (THS4511) chosen for its high bandwidth, low noise, and low distortion characteristics, is configured for a current amplification factor of 1500. Following the amplifier, a low-pass filter is implemented to reduce high-frequency noise and shape the signal. This shaping is necessary because the DAQ system, operating at a sampling frequency of \qty{40}{\mega\hertz} (\qty{25}{\nano\second}), must reliably capture single photoelectron pulses while preserving good energy resolution.
The low-pass filter has a cutoff frequency of \qty{2}{\mega\hertz}. Consequently, the bandwidth of the FEE (\qty{2}{\mega\hertz}) also acts as a shaping filter, extending the single photoelectron response time. The FEE also includes an offset adjustment feature that allows fine control over the baseline, preventing signal saturation and fully utilizing the dynamic range of the DAQ FPGA. Finally, a differential driver ensures sufficient current for transmission to the external acquisition system. Further details about readout system are covered in detail in Section~\ref{sec:daq}.

An LED calibration system is distributed along the EP plate in order for calibrating the TP at the opposite side of the detector. A set of 12 LEDs are distributed along the plane and placed inside the cavities of the EP (next to the PMTs), ensuring that the light goes through the sapphire window. 
Temperature sensors have also been installed on the EP. 

\section{Tracking Plane}\label{sec:tp}
The particles interacting in the NEXT-100 detector can be 3D reconstructed thanks to a Tracking Plane consisting of a matrix of 3,584 SiPMs that are placed in contact with the gas xenon (Fig.~\ref{fig:TP}). The sensors of \qtyproduct{1.3 x 1.3}{\milli\meter} (69\% larger area than the ones used in NEXT-White) effective photosensitive area are the S13372-1350TE model from Hamamatsu, with \qty{50}{\micro\meter} pixel pitch. These photo-sensors have been chosen due to their high sensitivity, and the reduced radioactivity in the optical detector package. The SiPMs are distributed on 56 Kapton PCBs (named as DICE-Boards) of {8$\times$8} SiPMs each at a pitch of \qty{15.55}{\milli\meter}. Compared to the SiPMs used in the NEXT-White detector, the density of SiPMs has been reduced. This was necessary due to the high density of outer cables, which were interfering with sensor connections. Simulations have demonstrated that, thanks to the high-quality reconstruction algorithms developed by the NEXT collaboration \cite{NEWTrack2}, reducing the SiPM coverage from 1.78$\%$ to 0.74$\%$ and distributing them uniformly along the TP does not impact the spatial resolution of the detector. This reduction cuts the number of channels by a factor of two.  

The DICE-Boards are placed at a distance of \qty{15.1}{\milli\meter} with respect to the anode mesh, and therefore at a distance of \qty{24.8}{\milli\meter} to the EL gate. The design concept used for the DICE boards is the same as the one for the NEXT-White detector: a low-radioactivity Kapton printed circuit of \qty{0.2}{\milli\meter} (Fig.~\ref{fig:TP_kapton}) with a flexible pigtail. 
As the glue in between different Kapton layers of these circuits was identified as the main source of radioactive impurities, a single layer of Kapton in the DICE-Boards where the SiPMs are soldered (the frontmost side of the circuit) was implemented. However, the flexible tail that crosses through the copper retained the extra layers of Kapton, to protect the lines from touching the copper as it is partially shielded from the detector by the ICS. 

The reflectivity of the detector and the light collection in the EP is maximized by placing PTFE masks of \qty{6}{\milli\meter} thickness (Fig.~\ref{fig:TP_TPB}) over the DICE-Boards. These masks have a symmetric hole of \qtyproduct{6 x 6}{\milli\meter} to ensure full light collection of the SiPMs at the TP. Both Kapton DICE-Boards (with mounted SiPMs) and PTFE masks are coated with a thin layer of TPB. Each of the DICE-Boards has a temperature sensor to monitor the system and also a blue LED to allow the calibration of the EP at the opposite side of the detector.

\begin{figure}[t!]
  \centering
    \includegraphics[width=0.8\linewidth]{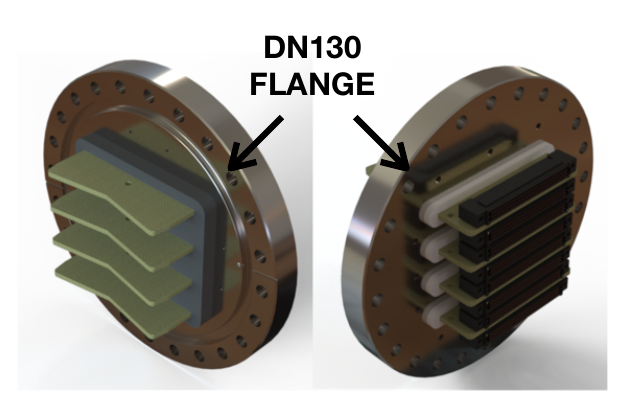}
    \includegraphics[width=0.8\linewidth]{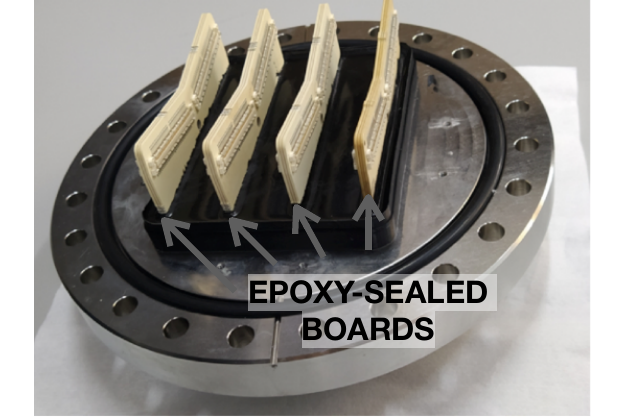}
    \caption{The NEXT-100 tracking plane feed-through. It consists of four epoxy-sealed boards mounted on a DN130 flange, each accommodating up to 16 DICE per feed-through.}
    \label{fig:TPFT}
\end{figure}

The flexible pigtails of the NEXT-100 DICE-board pass through the elongated slots of the ICS, where they connect to another Kapton cable that transmits the signal to a set of feed-throughs. A set of custom made NEXT-100 TP feed-throughs are designed for the experiment and represent a breakthrough that allow us to extract a very high number of channels from the high pressure vessel. Each of these uniquely designed feed-throughs consists of four epoxy-sealed boards mounted on a DN130 flange (Fig.~\ref{fig:TPFT}). Each board can accommodate four DICE boards, allowing a single feed-through to house up to 16 of them. This high-density, low-noise feed-through preserves the integrity of the SiPM signals and supports up to 2240 sensors per unit.\\
These feed-throughs have undergone rigorous mechanical validation, including pressure cycles, vacuum tightness tests with helium, and permeability tests with helium and neon. They have been tested for outgassing through residual gas analysis (RGA), which helped optimize the assembly process to avoid contamination from tapes or other consumables during epoxy application. The epoxy and electronic boards showed no apparent outgassing in RGA studies.\\
Additionally, the feed-throughs have passed X-ray inspections to verify the correct deposition of the epoxy, ensuring no bubbles or imperfections that could affect performance. Measurements indicate an exceptionally low helium leak rate of \qty{5.5e-10}{\milli\bar\cdot\liter\cdot\second^{-1}}. It has demonstrated the ability to withstand at least 200 pressurization cycles with a helium permeability of \qty{1e-6}{\milli\bar\cdot\liter\cdot\second^{-1}} at \qty{17}{\bar} and no detectable permeability for argon and neon. These tests confirm its robustness and reliability for demanding operational conditions.

\section{Read-out and DAQ}\label{sec:daq}


While the events of interest for the physics exploitation of NEXT-100 cover the $\sim${1 -- \qty{2.5}{\mega\electronvolt}} range with a trigger rate {$<$1~Hz}, calibration sources induce energy deposits from \qty{41.5}{\kilo\electronvolt} ($^{83m}$Kr) to \qty{2.6}{\mega\electronvolt} ($^{228}$Th) with trigger rates of tens of Hz. The acquisition system of NEXT-100 is able to cope with both data taking conditions. In particular, the DAQ implements a dual trigger scheme, with independent parameters, allowing for the continuous recording of Kr events (Type 1 trigger), meant to perform a volumetric calibration of the detector, while collecting also physics data (Type 2 trigger).

\begin{figure*}[t!]
  \centering
   \includegraphics[width=0.9\linewidth]{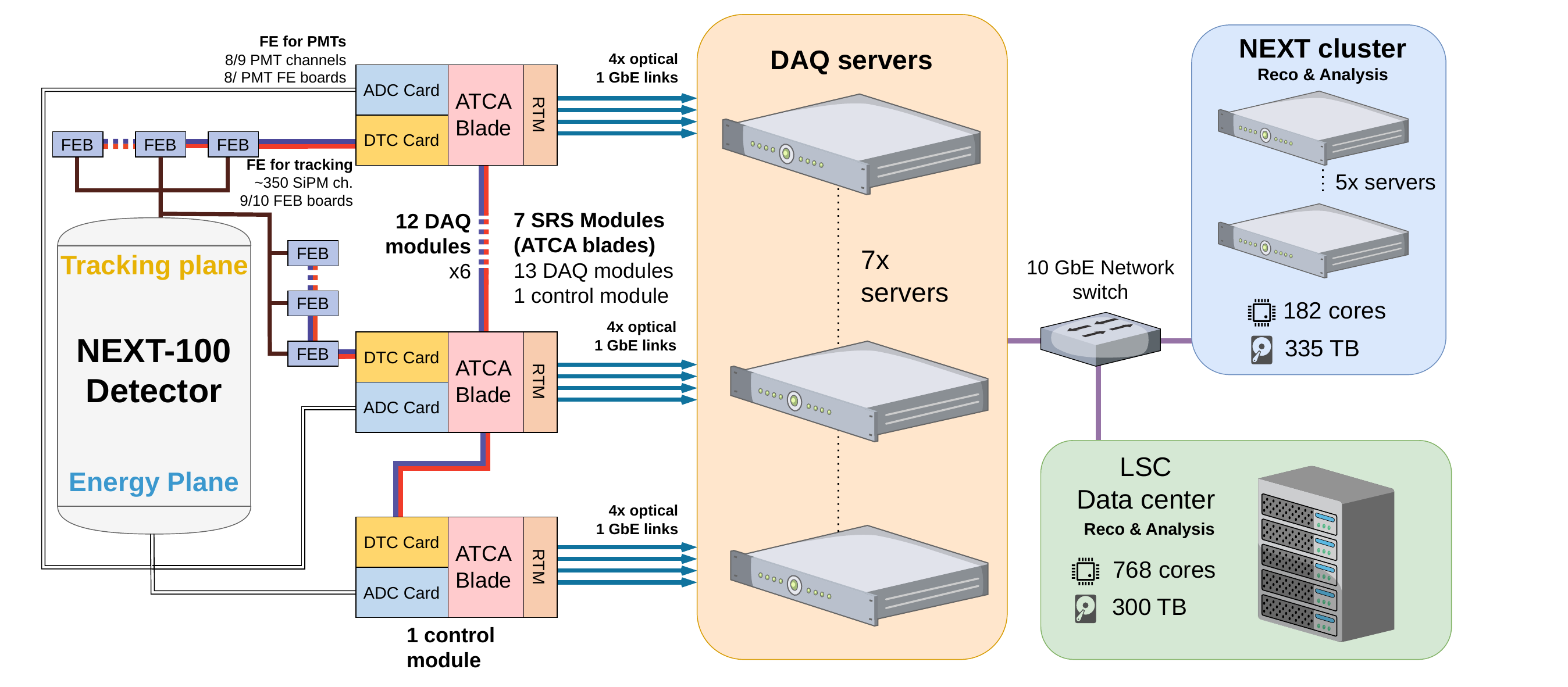}
  \caption{NEXT-100 Data Acquisition and Front-End hardware architecture. The system poses 13 DAQ modules reading the front-end electronics from both (TP and EP) sensors planes. The DAQ modules consist in ATCA blades that send data to the DAQ servers. The acquired data is reconstructed and analysed on two clusters: one from the NEXT collaboration (182 cores) and one from the LSC (768 cores).}
  \label{fig:daq}
\end{figure*}

\subsection{Read-out and DAQ Electronics}
The NEXT-100 experiment inherits the read-out and DAQ electronics, firmware, and software from its predecessor, the NEXT-White experiment \cite{NEWDAQ}. This reliable and well-established system was scaled for NEXT-100 by simply adding modules, eliminating the need for extra concentration stages. The read-out, DAQ and FEE architecture in NEXT-100 is represented in Fig.~\ref{fig:daq}.
The system has a total of 13 DAQ Modules, 7 more than its predecessor (divided into 7 EP DAQ Modules, and 6 TP DAQ Modules, reading out the 53 PMT and 3,584 SiPM sensors, respectively), and one extra Control Module for control and event detection. The main component in the DAQ system hardware is the advanced telecommunications computing architecture (ATCA) blades, a field programmable gate array (FPGA) based module originally developed for the CERN RD-51 Collaboration \cite{Readout}, that builds and buffers sub-events from incoming detector data as well as to perform system control. This module and the rest of the cards used in the DAQ (the two types of mezzanine, ADC and DTC, and the FEB cards used to readout SiPM sensors) are described in  \cite{NEWDAQ} and \cite{DAQFEB}. 

\subsection{DAQ Software}
Each half of an ATCA blade sends data to the DAQ servers via user datagram protocol (UDP) frames using two optical fiber gigabit Ethernet (GbE) links. In total, there are {28~GbE} links reaching the server farm. The data are read by 7 servers (in orange in Fig.~\ref{fig:daq}) running the data acquisition software, which has been developed from scratch and has been successfully tested and deployed in NEXT-100. The software's backend is written in Go \cite{GOSoftware}. It includes four main parts:

\begin{itemize}
    \item Local Data Concentrator (LDC): One program to readout the UDP frames from all the links connected to each search. There is one Go routine per link. Once all the fragments have been collected, a header is added and sent to the Global Data Concentrator.
    \item Global Data Concentrator (GDC): An event builder program. This program listens for Transmission Control Protocol (TCP) connections from LDCs containing subevents data. Once all the fragments of a given event are received, the whole event is put together with a new header and written locally to the output file.
    \item Server Application Programming Interface (API): The whole system requires a web server with a REST API to control the system and monitor its status. The API server communicates with the LDCs and GDCs using the gRPC protocol.
    \item Web Interface: The user interface is provided via a web page written with \textit{Vue.js} that communicates with the API web server, allowing the user to interact with the system.
\end{itemize}

Data received by the GDCs is written into binary files that are copied to the NEXT processing cluster (in blue in Fig.~\ref{fig:daq}) where they are decoded. Events on each file are reconstructed with the analysis software Invisible Cities \cite{NEXT_Collaboration_Invisible_Cities}. Some of the jobs are running in the LSC computing cluster (in green in Fig.~\ref{fig:daq}), that provides service to all LSC experiments. Custom software has been developed to manage productions and automatically send all required jobs. The DAQ servers, NEXT cluster and LSC cluster are connected via {10~GbE} network. The data production is monitored via Grafana dashboards.

\subsection{DAQ Event Detection}
The different topologies and orientation of the tracks produced in the active volume lead to a wide range of lengths in the drift axis (z) and thus in the time duration and height of the amplified signal. To deal with this variety of PMT pulse shapes, the NEXT-White detector was already equipped with specialized event detection capabilities \cite{EventDetect}, which have been also implemented in the NEXT-100 DAQ system, based on the charge, height and duration of the S2 signal in the PMTs. Although a S1-based trigger is possible, the regular operation of NEXT-100 relies on a S2-based trigger, as the amplified signals are easier to detect and differ depending on the topology of the event. Furthermore, coincident triggers across multiple PMT channels can be used to reduce potential biases in the event selection. In order to ensure the recording of the S1 signal associated to a S2 triggering the detector, as well as to guarantee that the full S2 signal is processed, the trigger readout window spans twice the maximum drift time, being the trigger time (start of the S2 signal) in the centre. 

Scaling from NEXT-White to NEXT-100 increased the number of sensors, detector volume, and buffer size, thereby raising the per-event data load and DAQ throughput requirements. Each of the seven servers can sustain up to 125~MB/s, for a combined limit of 875~MB/s, which is insufficient for uncompressed calibration runs. To address this, both lossy and lossless compression schemes were evaluated, with the adopted solution achieving a $\sim80\%$ reduction in data size for both PMTs and SiPMs \cite{DataComp}. This significantly lowers throughout demands and reduces DAQ dead time, keeping operation within design limits and avoiding costly system upgrades. 

As mentioned in the introduction to Section~\ref{sec:daq}, the continuous calibration with an internal $^{83\text{m}}$Kr source implies a Type 1 trigger rate of tens of Hz. This high rate induces a non-negligible dead time in the Type 2 trigger, meant to collect physics events, since both type of triggers share the double buffer and the Ethernet links used to transmit data to the DAQ servers, even if processed in parallel. To mitigate this, the NEXT-100 DAQ also employs a double-buffering strategy: Type 1 and Type 2 events are processed in parallel, with Type 1 events confined to a single buffer (with shortened acquisition window if necessary), ensuring that at least one buffer remains available for Type 2 events. With this configuration, initial tests suggest dead times from 11\% to 16\% for Type 2 trigger, being the double for Type 1. These dead times do not significantly impact the physics exploitation of NEXT-100 detector.

\begin{figure*}[h!]
    \centering
    \rotatebox{0}{\includegraphics[width=1.0\linewidth]{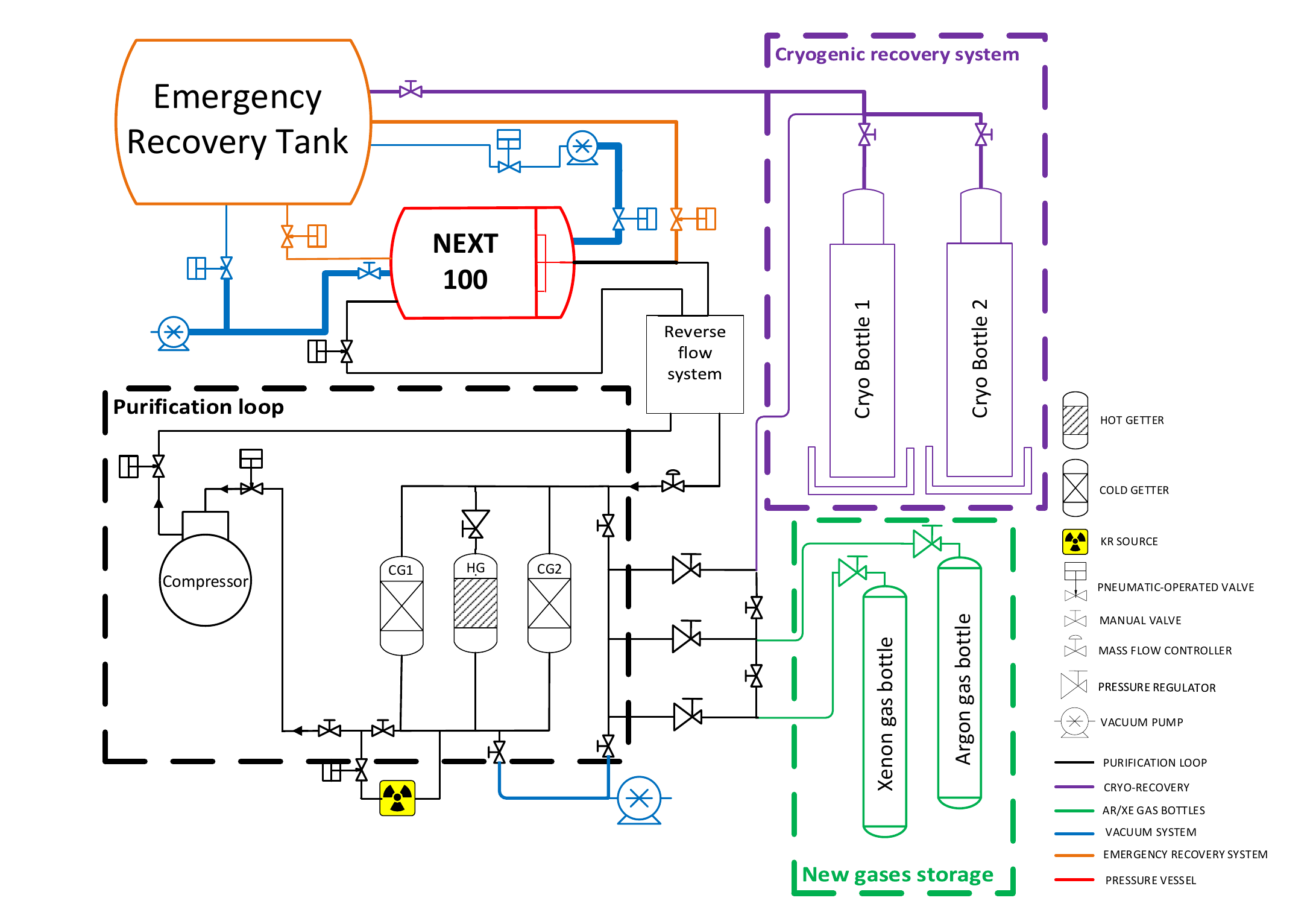}}
        \caption{Simplified representation of the gas system for the NEXT-100 detector. Highlighted with colors are the different parts of the system. }
    \label{fig:GS}
\end{figure*}

\section{Gas System}\label{sec:gas}
Ionization electrons created when a charged particle interacts in the detector can be lost due to attachment to electronegative impurities during drift towards the EL region. A low electron drift lifetime would imply a significant fraction of electrons lost before reaching the readout TP, resulting in signal degradation, inaccurate energy reconstruction, and poor spatial resolution. A good control of the gas impurities is needed to ensure a constant high electron lifetime. During the design phase of the NEXT-100 detector, a careful testing for outgassing of the different detector materials in contact with the gas was performed, also using the experience of the NEXT-White detector operation. 
During detector operation, the ultimate goal of the NEXT-100 gas system is to monitor and maintain gas purity under control, and keep the detector at a stable pressure. \\

The NEXT-100 gas system builds upon the former NEXT-White \cite{NEWDetector} one and can be partitioned into four interconnected subsystems, pictured in Fig.~\ref{fig:GS}: 
\begin{itemize}
    \item[(a)] The \textit{pressure system} (red and black lines), encompassing gas recirculation and purification. A compressor is used to keep the pressure of the system stable at the operating value and to circulate the gas. The detector pressure is monitored at different points allowing the observation of small variations due to temperature oscillations (very important due to the size of the NEXT-100 detector). Purification of the gas is achieved by circulation at around $\sim$100~slpm through chemical purifiers: two ambient temperature \textit{cold getters} (CG1 and CG2), and one \textit{hot getter} (HG). The maximum operation pressure of the NEXT-100 detector is \qty{13.5}{\bar}. This value is limited by the rupture discs of the recovery system, which reduce the maximum operating pressure of the vessel (\qty{15}{\bar}). 
    \item[(b)] The \textit{vacuum system}, including the vacuum lines (blue) and pumps. 
    The vacuum system is used to evacuate the system prior to its filling with xenon. This helps to considerably reduce the time needed to purify the gas. The NEXT-100 vacuum system is able to hold a vacuum of \qty{1e-6}{\milli\bar}. 
    This system has been considerably upgraded from the NEXT-White one by increasing the size of the main lines, using rigid pipes and by multiplying the number and positions of pump stations (combination of a turbomolecular pump with a roughing pump). This allows reaching a good vacuum level in such a large system within an acceptable amount of time.
    \item[(c)] The \textit{emergency recovery system} (orange lines) of which the main part is the expansion tank. Thanks to the constant monitoring of the pressure and vacuum values at the gas system, the system is able to react quickly and without any manual intervention to changes indicating an unexpected overpressure condition or a potential leak. The emergency recovery system is designed to collect the maximum gas capacity of the pressure vessel on a recovery tank in \qty{11}{\minute}. 
    \item[(d)] The \textit{cryo-recovery system} (purple lines), including the cryogenic recovery bottle and all the lines connecting to the rest of the system. The recovery of the xenon under normal operation conditions is achieved by the connection of two cryo-bottles where the xenon is collected by freezing it, to the recirculation loop and to the expansion tank (one of the bottles is used to recover normal xenon, while the other is for enriched xenon). 
\end{itemize}

Major upgrades have been implemented in the gas system considering the bigger size of the NEXT-100 detector and lessons learned with NEXT-White. First, a metering valve to improve the regulation of the krypton input to the gas system has been installed (represented with radioactive symbol on Fig.~\ref{fig:GS}). The remotely controlled metering valve facilitates krypton source survey operations, enabling fine adjustments to achieve the desired number of krypton events inside the NEXT-100 detector. Additionally, some major updates regarding the emergency recovery and the circulation flow have been implemented and are covered in the next subsections. 

\begin{figure*}
    \centering
    \begin{subfigure}{0.40\textwidth}
        \centering
        \rotatebox{270}{\includegraphics[width=0.80\linewidth]{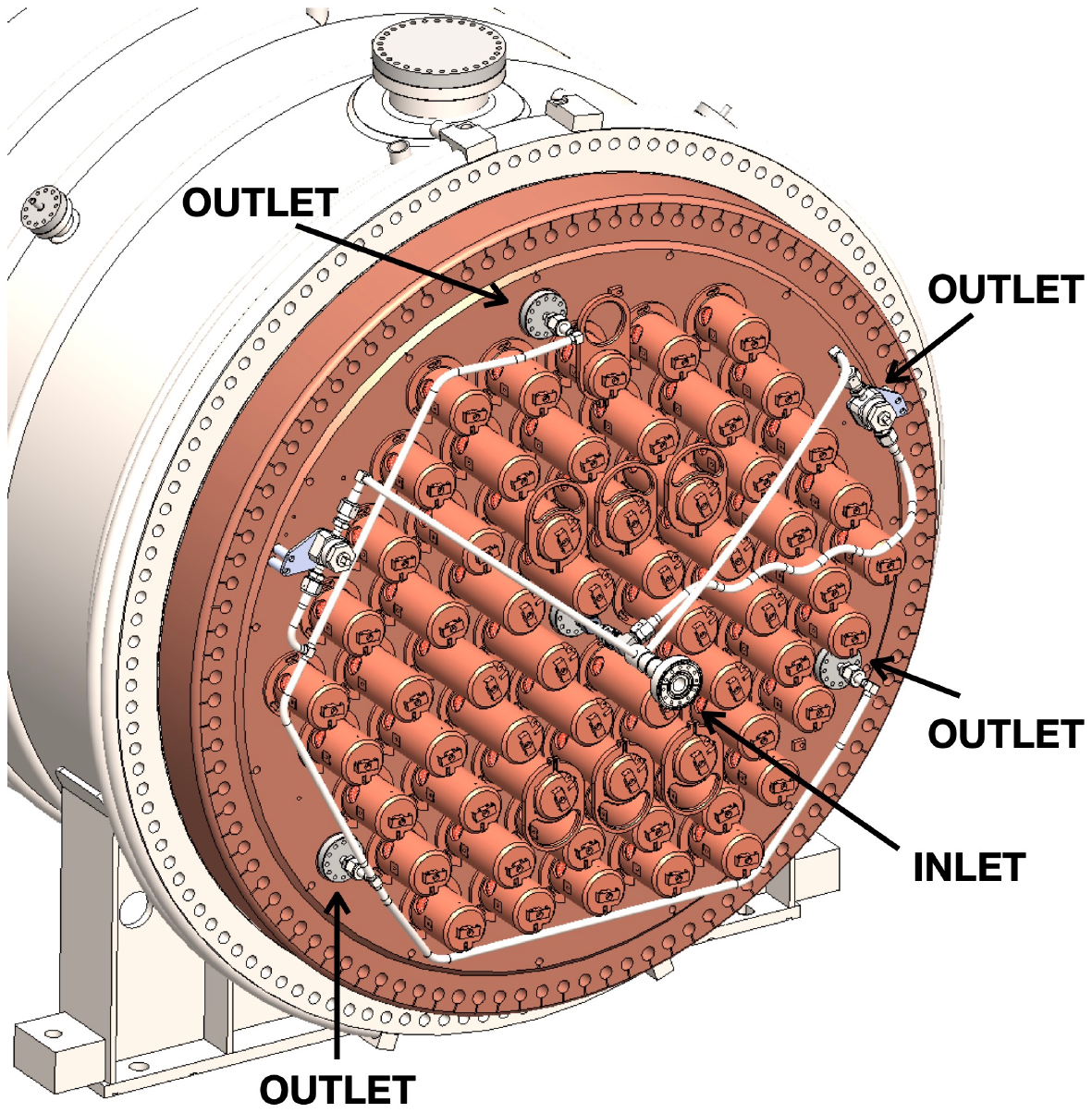}}
        \caption{}
        \label{fig:GAS_Flow_Pipes}
    \end{subfigure}
    \begin{subfigure}{0.50\textwidth}
        \centering
        \includegraphics[width=1\linewidth]{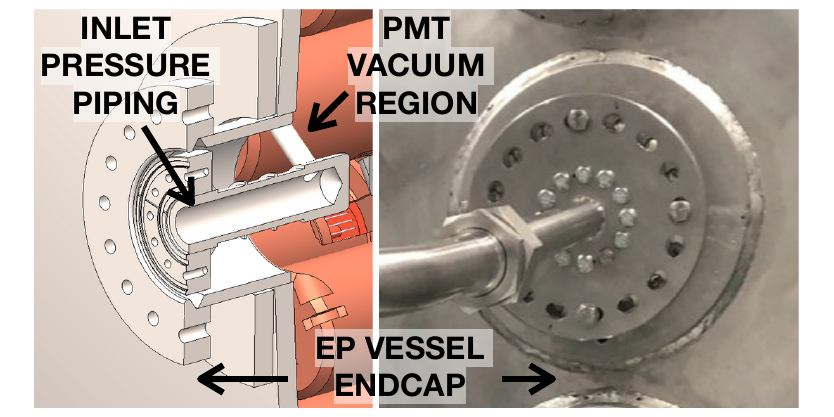}
        \caption{}
        \label{fig:GAS_Inlet}
    \end{subfigure}\\
    \begin{subfigure}{0.45\textwidth}
        \centering
        \rotatebox{0}{\includegraphics[width=1.\linewidth]{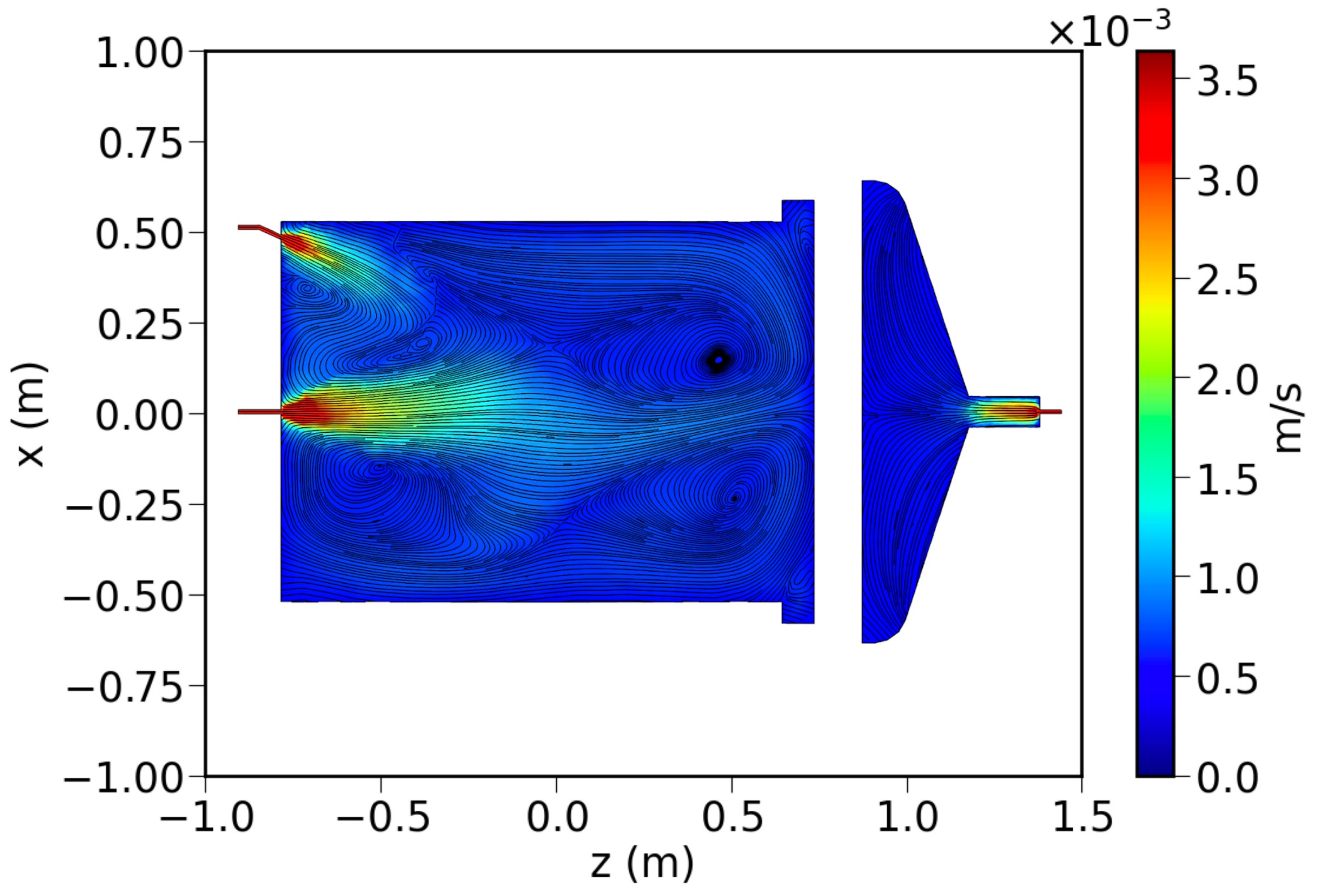}}
        \caption{}
        \label{fig:GAS_Flow_EP}
    \end{subfigure}
    \begin{subfigure}{0.45\textwidth}
        \centering
        \rotatebox{270}{\includegraphics[width=0.70\linewidth]{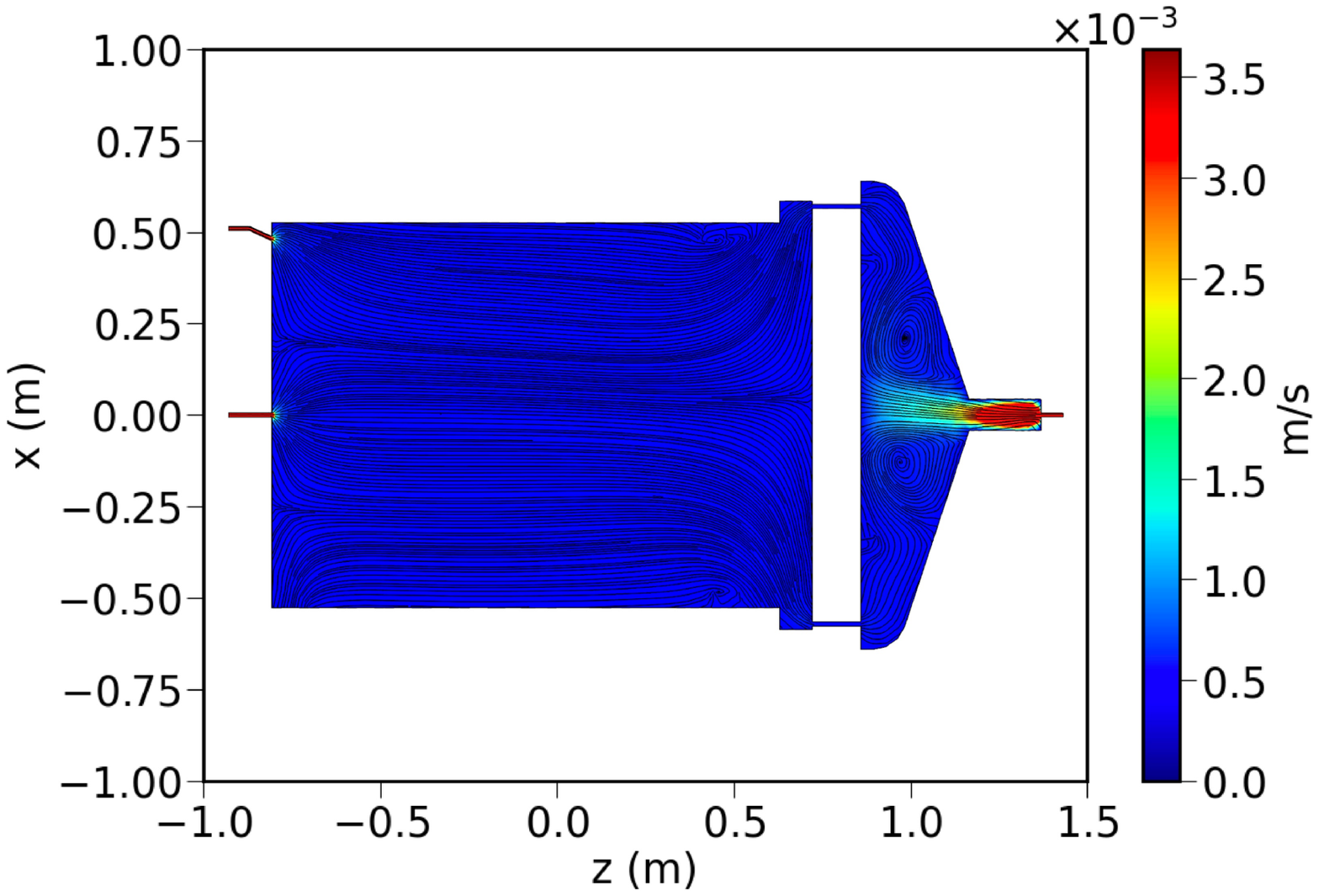}}
        \caption{}
        \label{fig:GAS_Flow_TP}
    \end{subfigure}
    \caption{(a) Gas injection system placed behind the Energy Plane copper shielding to improve circulation flow. (b) Port at the energy plane cap that controls the gas flow of the injection system in the PMT vacuum region. Simulations of the gas flow when the circulation is performed from (c) Energy Plane towards Tracking Plane and (d) Tracking Plane towards Energy Plane. The implementation of a reverse flow configuration provides the capability to study the influence of flow patterns within the system. Simulations provide a homogeneous flow on both directions. }
    \label{fig:GAS_Flow}
\end{figure*}

\subsection{Emergency Recovery}
The recovery system of the NEXT-100 experiment (orange line in Fig.~\ref{fig:GS}) has undergone a significant upgrade with the addition of a newly manufactured recovery tank. This tank has a capacity of \qty{20}{\meter^{3}} and is rated for a maximum pressure of \qty{3}{\bar}. It has demonstrated the ability to maintain impurity levels below \qty{1}{ppm} over extended periods (several months) without the need for continuous evacuation. This performance is attributed to its electropolished internal surface, achieving a roughness below \qty{0.6}{\micro\meter}, and a high-temperature nitrogen gas treatment at \qty{200}{\celsius} applied during manufacturing. These features significantly reduce maintenance requirements while ensuring long-term xenon purity. The tank includes a dual-stage overpressure protection system. The first stage, newly introduced in this upgrade, consists of a rupture disc rated at \qty{2}{bar}, a relief valve also set at \qty{2}{bar}, and a manual valve for maintenance. This is designed to protect the emergency recovery tank to avoid a loss of gas in case of overheating or fire at the laboratory. The second stage, retained from the previous setup, consists of a rupture disc rated at \qty{3}{bar}, which protects the maximum operating pressure of the emergency recovery tank. These two stages provide redundant safety mechanisms, ensuring robust protection during overpressure events. Unlike the previous system, this upgraded configuration does not require continuous vacuuming; instead, vacuum cycles can be scheduled at controlled intervals, improving operational stability and enhancing safety in xenon recovery and storage. 

One possible critical failure in the NEXT-100 detector is the potential rupture of a PMT window. The PMTs used are not designed to operate under pressure. To address this, custom windows were developed to house the PMTs in a separate vacuum environment. These windows undergo rigorous validation procedures, including gas permeability tests, helium leak testing under both vacuum and pressure, and pressure cycling, to ensure reliability. Nevertheless, a window failure could result in xenon leakage and pressure damage to the whole EP. To mitigate this risk, the EP is directly connected to the recovery tank via a turbo pump (dark blue line in Fig.~\ref{fig:GS}). The large volume of the recovery tank allows it to function as a primary buffer, offering a direct path for gas evacuation in case of a window breach. Simulations have verified that this setup keeps the EP pressure below \qty{3}{\bar} even in the event of a catastrophe window failure, minimizing potential damage to the PMTs. Additionally, this configuration enables continuous xenon monitoring with an RGA and rapid leak detection, thereby improving the overall safety and responsiveness of the system. 



\subsection{Improvement of the circulation flow}

To further enhance the gas purification process and electron lifetime uniformity along the NEXT-100 detector a series of gas flow simulations were performed. The most optimal solution was the introduction of gas input holes in the EP copper plate, with pipes going through the vacuum volume (Fig.~\ref{fig:GAS_Flow_Pipes}). This results in a uniform distribution of the gas along the active region of the detector, with fewer dead regions than what we would have had with a single pipe in the center of the vessel. The system also allows for selective control of the gas flow across different regions of the EP, allowing it to be directed towards the peripheral section, the central section, or both simultaneously. 
Additionally, two manual valves provide the ability to select the flow path (between outlet holes and/or the inlet hole) and to isolate the pressure volume. The first function makes it easier to clean areas of the detector where gas circulation is reduced. The second function enables the EP endcap to be opened by removing only a small section of the pressure piping. This is achieved by routing the outlet through a dedicated port and employing a secure, sealed connection system (Fig.~\ref{fig:GAS_Inlet}). This connection features blind-threaded holes that allow it to be tightened from the ambient pressure side, so the EP endcap can be opened while keeping the internal pressure piping securely in place. Simulations of the gas flow were performed and are represented in Fig.~\ref{fig:GAS_Flow_EP} and Fig.~\ref{fig:GAS_Flow_TP}. In addition, a modification has been introduced to the gas system, allowing simultaneous operation of both cold getters along with the hot getter. This upgrade provides a range of configurations that significantly increase the versatility of the purification circuit.

\section{Radiopurity}\label{sec:radiopurity}

\begin{table*}[]
\centering
\caption{Summary of screening measurements for in-vessel materials and components near the energy endcap, tracking endcap, and barrel detector regions. Energy end-cap covers cathode and PMTs, tracking end-cap covers gate, anode and SiPMs, and barrel region covers the TPC and the ICS volumes. For each region, components are listed in decreasing order of total activity (\Bi{214} plus \Tl{208}). Only materials or components contributing more than 1~mBq are separately listed, with less active contributions grouped under “other” components. Components that do not comply with our radiopurity requirements ($>$0.1count/year after selection cuts) are highlighted in italic.}
\label{table:radiopurity}
\begin{tabular}{m{0.5cm}|p{5cm}|p{2.3cm}|c|c|rr}
\hline
 & \textbf{Material or component} & \textbf{Provider} & \textbf{Quantity} & \textbf{Technique} & \multicolumn{2}{c}{\textbf{Total activities} [mBq]}  \\
& &  &  & & \Bi{214} & \Tl{208} \\ \hline
\multirow{10}{*}{\rotatebox{90}{ENERGY END-CAP}} &SS316Ti M18 screws for copper plate &  & 3.6E+01~kg & Ge & \textit{8.0E+01} & \textit{4.7E+01}  \\
&Sapphire windows with PEDOT coating and copper cans &  & 6.0E+01~units & Ge & \textit{$<$6.6E+01} & \textit{$<$2.2E+01} \\ 
&PMT bases &  & 6.0E+01~units & Ge & \textit{4.1E+01} & \textit{1.5E+01} \\
&PMTs & Hamamatsu & 6.0E+01~units & Ge & 2.1E+01 & 1.1E+01 \\
&Gel for window-PMT optical coupling & NuSil & 2.4E-01~kg & Ge & $<$7.9E+00 & $<$3.2E+00 \\
&Oxygen-free Cu for EP plate & Atlantic Copper & 1.4E+03~kg & ICP-MS & 1.6E+00 & 1.4E-01 \\

& Stainless steel cathode mesh & & 1.7E-01~kg & ICP-MS & 1.3E+00 & 5.0E-02 \\
& Other &  &  &  & $<$1.1E+00 & $<$3.7E-01 \\
& & & & &  &\\
& {\bf Sub-total} & & & & {\bf $<$2.2E+02} & {\bf $<$9.9E+01} \\
\hline 
\vspace{0.2cm}

\multirow{8}{*}{\rotatebox{90}{TRACKING END-CAP}} & Kapton cable connectors FX11LA & Hirose & 5.6E+00~units & Ge & \textit{1.6E+02} & \textit{1.2E+02} \\
& PTFE masks & FEYMA & 5.6E+00~units & Ge & \textit{$<$1.4E+01} & $<$4.2E+00 \\
& SiPMs S13372-1350TE & Hamamatsu & 3.6E+03~units & Ge & \textit{$<$1.2E+01} & $<$2.8E+00 \\
& Kapton DICE-Boards &  & 5.6E+00~units & Ge & 3.9E+00 & $<$5.8E-01 \\
& Stainless steel gate and anode meshes & & 3.4E-01~kg & ICP-MS & 2.5E+00 & 1.0E-01 \\
& Oxygen-free Cu for TP plate & Atlantic Copper & 1.5E+03~kg & ICP-MS & 1.8E+00 & 1.5E-01 \\
& Si-bronze for gate and anode rings & Farmers Copper & 1.7E+01~kg & ICP-MS & 1.1E+00 & 2.8E-01  \\
& Other &  &  &  & $<$1.2E+00 & $<$3.2E-01 \\
& & & & &  &\\
& {\bf Sub-total} & & & & {\bf $<$1.9E+02} & {\bf $<$1.3E+02} \\
\hline \vspace{0.2cm}

\multirow{10}{*}{\rotatebox{90}{BARREL REGION}} & Electrolytic Copper for inner shield and field rings & Welding Copper & 6.1E+03~kg & ICP-MS & $<$4.5E+01 & $<$7.1E+00 \\
& Field cage resistors & Vishay Techno & 1.6E+02~units & Ge & \textit{2.7E+01} & 2.5E+00 \\
& SS316Ti M6x20 screws for inner copper shield &  & 7.2E-01~kg & GDMS & 8.9E+00 & 7.4E-01  \\
& PTFE reflector panels & PTFE Cube & 5.5E+01~kg & ICP-MS & 2.8E+00 & 2.9E-01 \\
& HDPE wrap & Plastenics & 5.0E+01~kg & ICP-MS & 1.8E+00 & 1.1E+00 \\
& HDPE struts & Plastenics & 3.9E+01~kg & ICP-MS & 1.4E+00 & 8.5E-01 \\
& Other &  &  &  & $<$1.1E+00 & $<$2.8E-01 \\
& & & & &  &\\
& {\bf Sub-total} & & & & {\bf $<$8.8E+01} & {\bf $<$1.3E+01} \\
\hline \vspace{0.2cm}
& {\bf Total} & & & & {\bf $<$5.0E+02} & {\bf $<$2.4E+02} \\
\end{tabular}
\end{table*}

NEXT-100 sensitivity studies assume background levels of order \qty{4e-4}{\text{counts}\per(\kilo\electronvolt\cdot\kilogram\cdot\text{year})} after all \lessnuBB\ selection cuts~\cite{NEXT:2015wlq}, or about $B_{tot}=1$~counts/year total background rate in the entire energy region of interest, for energy resolution and xenon mass values characteristic of NEXT-100. As the background model accounts for many sources, ideally we set a requirement of less than 10\% of this total rate from any given (isotope, detector volume) background source combination $(i,j)$, that is $B_{i,j}=0.1$~counts/year. This background rate contribution is given by 
\begin{equation*}
B_{i,j} = (3.17\cdot10^{-5})\cdot A_{i,j}\cdot\epsilon_{i,j}\quad, 
\end{equation*}
where $A_{i,j}$ is the radio-activity (in mBq) in component $j$ produced by isotope $i$,  $\epsilon_{i,j}$ is the (dimensionless) background acceptance, that is the probability for a decay $i$ in detector component $j$ to induce an event in the detector surviving the \lessnuBB\ selection cuts, and $(3.17\cdot10^{-5})$ is a factor to convert mBq rate units into year$^{-1}$ units. Through MC simulations with a detailed detector geometry modelling and an idealized detector response and event reconstruction, we estimated the background acceptance factors $\epsilon_{i,j}$ for various (isotope, detector component) combinations and for a simple cut-based \lessnuBB\ event selection, as done in \cite{NEXT:2015wlq}. In this way, we can estimate maximum tolerable specific activities $a_{i,j}$ for each candidate material, where $a_{i,j}=A_{i,j}/M_j$, and $M_j$ is the mass of detector component $j$, providing an approximate guidance before accepting/rejecting any given material or component screened. For consistency, the same methodology described in \cite{NEXT:2015wlq} to define radiopurity requirements was kept throughout the several years that the material screening campaign for NEXT-100 lasted.

The NEXT collaboration, with the support of LSC, PNNL, SURF and external companies, has performed a comprehensive radiopurity screening campaign of all relevant materials and components installed in the NEXT-100 detector and surrounding infrastructures. Results of this campaign were used both for selecting materials and components prior to their manufacturing and installation in the detector, and are being used to build the NEXT-100 background model predictions. Particularly relevant is the screening of impurities in the \U{238} and \Th{232} decay chains, given that their \Bi{214} and \Tl{208} daughters are expected to be the dominant radiogenic background in the energy region of interest for \lessnuBB\ searches. 

Several techniques have been applied to quantify ultra-low levels of specific activities~\cite{Cebrian:2024avr}. Measurements based on gamma-ray spectroscopy with ultra-low background, high-purity, Ge detectors of the LSC Radiopurity Service have been performed on essentially all NEXT-100 materials and components. Being a non-destructive technique, Ge spectroscopy is particularly useful for detector assemblies involving many components, such as photo-sensors (PMTs, SiPMs) or electronic readout parts (PMT bases, Kapton DICE-Boards, connectors). For certain materials, more sensitive, but destructive, techniques have to be used. Glow discharge mass spectrometry (GDMS) or inductively coupled plasma mass spectrometry (ICP-MS) techniques provide concentrations of \U{238} and \Th{232} in material samples, and hence their activity. Assuming secular equilibrium, specific activities for \Bi{214} and \Tl{208} can be derived. GDMS is suitable only for metals. 

Table~\ref{table:radiopurity} summarizes central values and upper limits (95\% C.L.) for the \Bi{214} and \Tl{208} total activities of several detector components and materials installed within the NEXT-100 vessel. Quantities for each component or material are also reported, such that specific activities can also be evaluated. The table covers the information of the energy, tracking and barrel regions. For clarity, only total activities exceeding 1~mBq (\Bi{214} plus \Tl{208}) are shown. Overall, all in-vessel components are expected to contribute less than 1~Bq (\Bi{214} plus \Tl{208}), with comparable contributions from the three detector regions. Many more components belonging to the pressure vessel itself, or to the shielding structure and infrastructures located outside it, were also screened before being selected. They are not reported here for conciseness, as they tend to be less critical from a radiopurity point of view, given their larger (and shielded) distance from the detector active volume.

Screening measurements for most detector components and materials installed satisfy our radiopurity requirements. The few components that may not (note that only upper limits were obtained for certain specific activity measurements) are still estimated to be within a factor of 2--3 from requirements, at most. The only notable exception we are aware of are the Kapton cable connectors FX11LA from Hirose. Despite being shielded by copper, their activity is large enough that they surpass requirements by about one order of magnitude. Once suitable alternative connectors are identified, they will be replaced in a future detector intervention. Of course, higher background rates compared to our screening estimates may be observed in the future during NEXT-100 low-background operations, pointing to either unscreened contributions, or contamination of some component subsequent to its screening (e.g., during the manufacturing or installation process). Hence, a detailed background measurement analysis will be essential, as done for the NEXT-White detector~\cite{NEWBkg2}. As discussed in Ref.~\cite{NEXT:2018zho}, detector backgrounds induced by internal $^{222}$Rn activity within the xenon active volume are expected to be rejected very efficiently, and to contribute {0.1~counts/year} at most to the total NEXT-100 background budget after all selection cuts.

Prior to installation in the detector, all detector components and materials underwent a standardized cleaning procedure. This applies both to metallic (e.g., steel alloys, bronze alloys, copper, lead) and plastic (e.g., HDPE, PTFE) components. For copper pieces, the outer layers were first removed by manual wiping or scouring. The main cleaning step involves a soap cleaning solution with Alconox8 detergent in an ultrasonic bath, followed by rinsing with distilled water. Additional cleaning steps (nitric acid solutions, citric acid solutions, ethanol or isopropanol) were used for certain materials. All pieces were dried with nitrogen gas to remove water impurities, and stored in plastic bags until installation. Installation of all components was performed in controlled environmental conditions, by installing an ISO-7 clean room of 46~m$^2$ surface and 4~m height, fully enclosing the NEXT seismic table, the NEXT-100 pressure vessel and the outer lead shielding structure.

\section{Detector Operation}\label{sec:commissioning}
The NEXT-100 detector began operation in May 2024 and has undergone commissioning in two phases. A first campaign was conducted with argon gas at $\sim$\qty{4}{\bar}. The goal of this campaign was to test all the systems (energy and tracking  planes, gas system, DAQ, and HV for drift and EL fields) under regular operation conditions, and in particular to identify possible gas leaks in the vessel or the gas system. Once all systems were demonstrated to be leak-tight, a second commissioning stage started in October 2024 using $^{136}$Xe-depleted xenon gas at the same pressure. During both phases, PMTs and SiPMs were calibrated on a weekly basis in order to ensure stability and to define the long-term calibration strategy.
During the commissioning, both argon and xenon have been circulated through the CG, allowing for a quick purification of the gas. The cold getter produces a significant amount of $^{222}$Rn in the system (see \cite{NEWBkg1}). The alpha decays of this isotope and its progeny ($^{218}$Po and $^{214}$Po) have been used to characterize the response of the detector, without putting at risk the future background measurements. The detector operation conditions of both commissioning runs are summarized in Table~\ref{table:run_pars} together with estimated drift conditions.

\begin{table}[t!]
\centering
\caption{Configuration parameters of the commissioning runs of the NEXT-100 operation using argon and xenon gas. Conditions are different due to the differences in breakdown voltages of the two gases. Drift velocity and diffusion parameters (longitudinal and transversal, $D_{L}$ and $D_{T}$) are estimated with Magboltz \cite{Magboltz} using the operation parameters.}
\label{table:run_pars}
\begin{tabular}{l|ll}
\hline\noalign{\smallskip}
\textbf{Parameter} & \textbf{Argon} & \textbf{Xenon}\\
\noalign{\smallskip}\hline\noalign{\smallskip}
Pressure [bar] & $\sim$4 &  $\sim$4 \\
$V_{\mathrm{cathode}}$ [V] & 14,700 & 23,000\\
$V_{\mathrm{gate}}$ [V] & 6,700 & 9,000 \\
Detector Diameter [mm] & \multicolumn{2}{c}{1014} \\
Drift Length [mm] & \multicolumn{2}{c}{1187}\\
Drift Field [V/cm] & 74 & 118 \\
E/P [kV/cm/bar] & $\sim$ 1.47 & $\sim$ 2.30 \\
$v_{d}$ [mm/\si{\micro\second}] & 1.51 & 0.82 \\
$D_{T}$ [\si{\micro\meter}/$\sqrt{\text{cm}}$] & 1930 & 2040 \\
$D_{L}$ [\si{\micro\meter}/$\sqrt{\text{cm}}$] & 660 & 580 \\
\noalign{\smallskip}\hline
\end{tabular}
\end{table}

\begin{figure}
    \centering
    \begin{subfigure}{0.45\textwidth}
        \centering
        \rotatebox{0}{\includegraphics[width=0.85\linewidth]{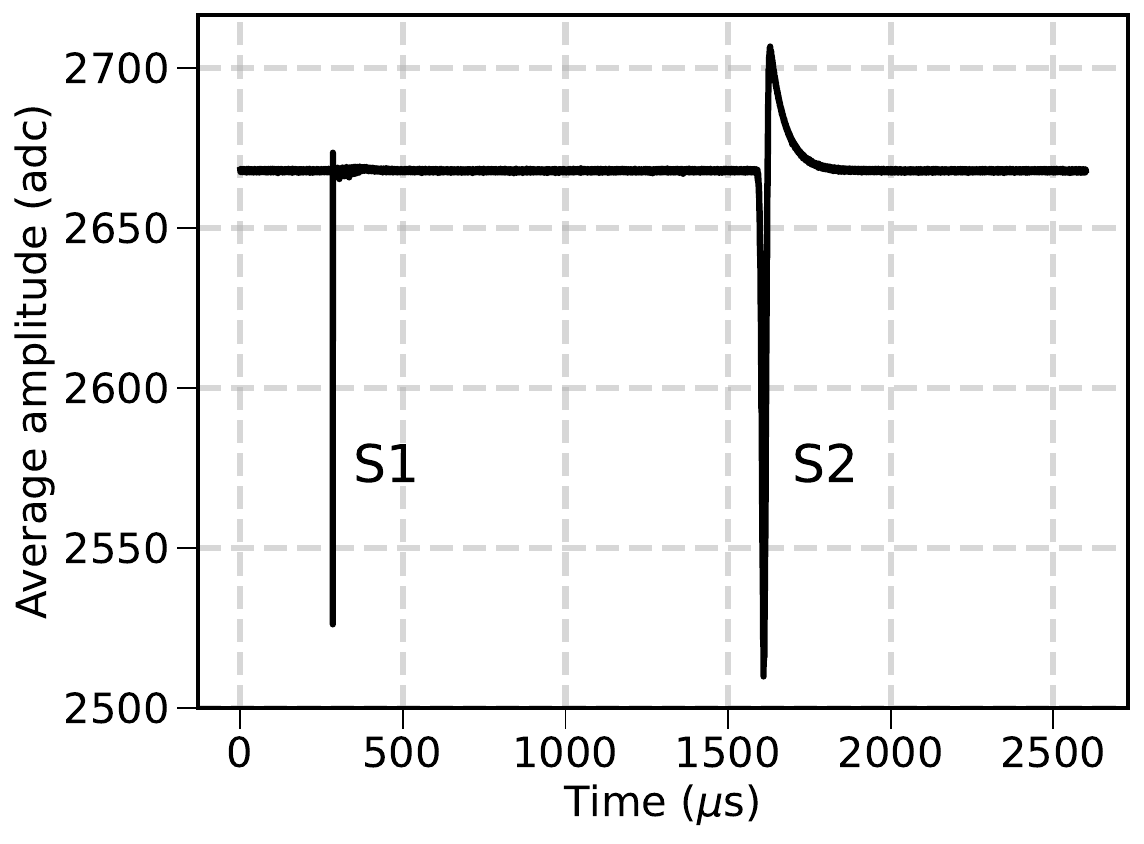}}
        \caption{}
        \label{fig:RWF}
    \end{subfigure}
    \begin{subfigure}{0.45\textwidth}
        \centering
        \rotatebox{0}{\includegraphics[width=0.9\linewidth]{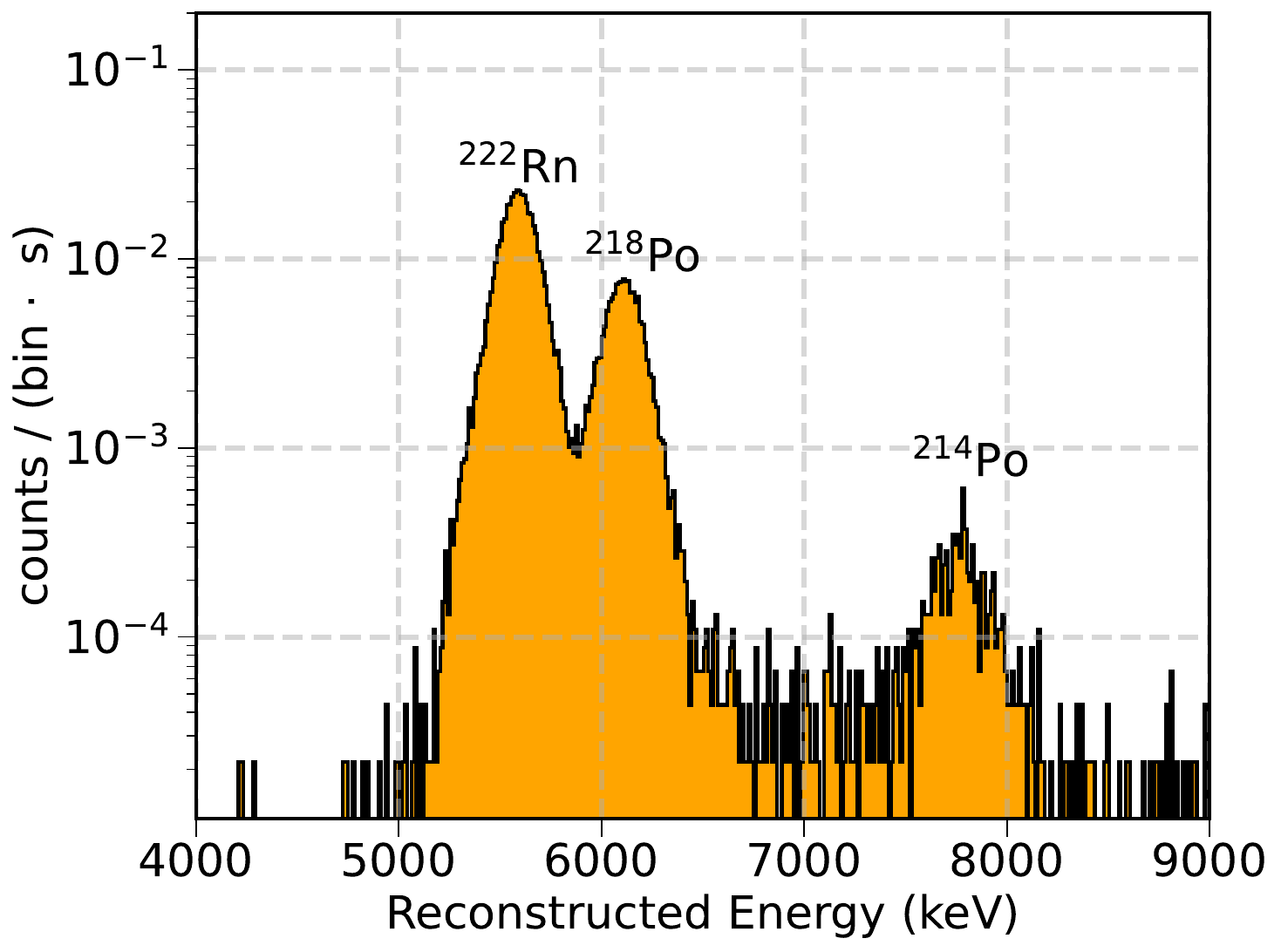}}
        \caption{}
        \label{fig:alpha_energy}
    \end{subfigure}
    \begin{subfigure}{0.45\textwidth}
        \centering
        \rotatebox{0}{\includegraphics[width=0.9\linewidth]{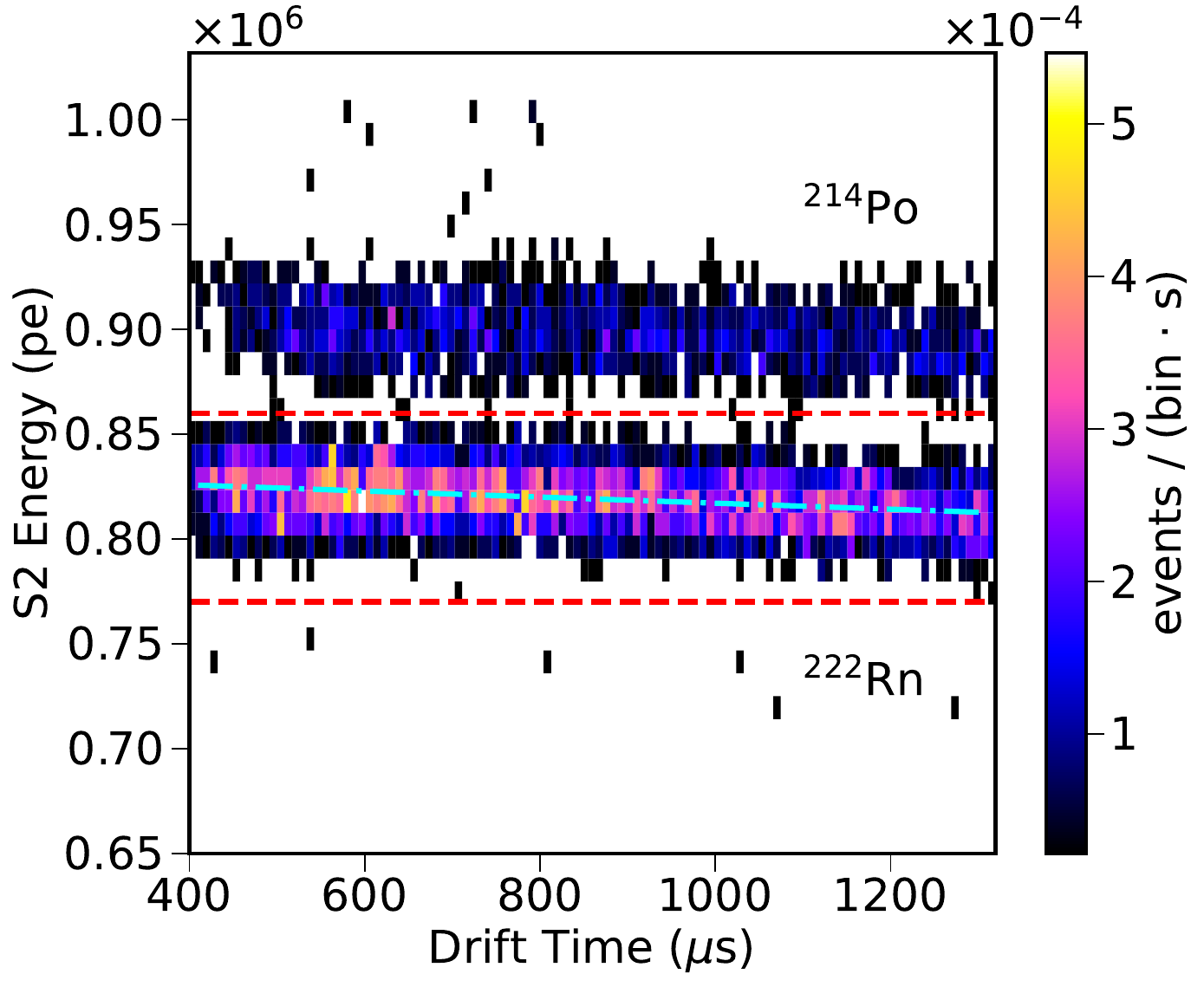}}
        \caption{}
        \label{fig:lifetime}
    \end{subfigure}
    \caption{(a) Average of the PMT waveforms in the NEXT-100 detector for an alpha candidate event. (b) Reconstructed energy of Rn-induced alpha decays in Xe: $^{222}$Rn (5.59 MeV), $^{218}$Po (6.11 MeV) and $^{214}$Po (7.83 MeV). The energy has been corrected by the electron lifetime. (c) S2 energy of the $^{222}$Rn and $^{214}$Po alpha decays in Xe as a function of the drift time. The dashed blue line shows the fit yielding a lifetime value of \SI{57(3)}{\milli\second}, where the uncertainty accounts only for the statistics in the data sample considered. The dashed red lines mark the energy range considered for the fit, covering only the alpha decays from $^{222}$Rn. The data correspond to a 24-hour period during the Xe commissioning phase.}
    \label{fig:alphas}
\end{figure}

\subsection{Commissioning with argon (May – Sept. 2024)}

During this period, the pressure conditions and performance of the gas system were monitored by means of the slow controls of the detector. The stability at high voltages applied at the cathode and the EL regions was also successfully tested by gradually increasing the fields. A drift field scan was performed varying the HV at the cathode from \qtyrange{13.0}{14.7}{\kilo\volt}, for a fixed voltage of \qty{6.0}{\kilo\volt} at the gate. This corresponded to drift fields (drift velocities) from $\sim$\qtyrange{60}{73}{\volt/\centi\meter} ($\sim$\qtyrange{1.4}{1.5}{\milli\meter/\micro\second}). An EL field scan was also conducted by increasing the HV at the gate from \qtyrange{6}{6.7}{\kilo\volt}, corresponding to reduced electric fields between $\sim$\qty{1.52}{\kilo\volt/\centi\meter/\bar} and $\sim$\qty{1.66}{\kilo\volt/\centi\meter/\bar}, while keeping the drift field constant. The operation voltages for this argon run were finally set to \qty{14.7}{\kilo\volt} at the cathode and \qty{6.7}{\kilo\volt} for the gate, values limited by the breakdown voltage in argon (\qty{6.93}{\kilo\volt}) \cite{Breakdown}. These conditions remained stable during the argon commissioning period, without the appearance of sparks or discharges. 

A DAQ trigger was configured to collect events above $\sim$1 MeV, relying on the S2 signals detected by one of the central PMTs. A typical raw waveform from an alpha candidate event is shown in
Fig.~\ref{fig:RWF}. The analysis of the Rn-induced alpha decays, uniformly distributed across the active volume, allowed for the characterization of the detector performance and its time stability. In particular, stable values for the electron lifetime and the drift velocity ($\sim$\qty{1.5}{\milli\meter/\micro\second}) were derived on a daily basis. The light yield was monitored, exhibiting only sub-percent variations. 

 \begin{figure}
    \centering
    \begin{subfigure}{0.45\textwidth}
        \centering
        \rotatebox{0}{\includegraphics[width=0.95\linewidth]{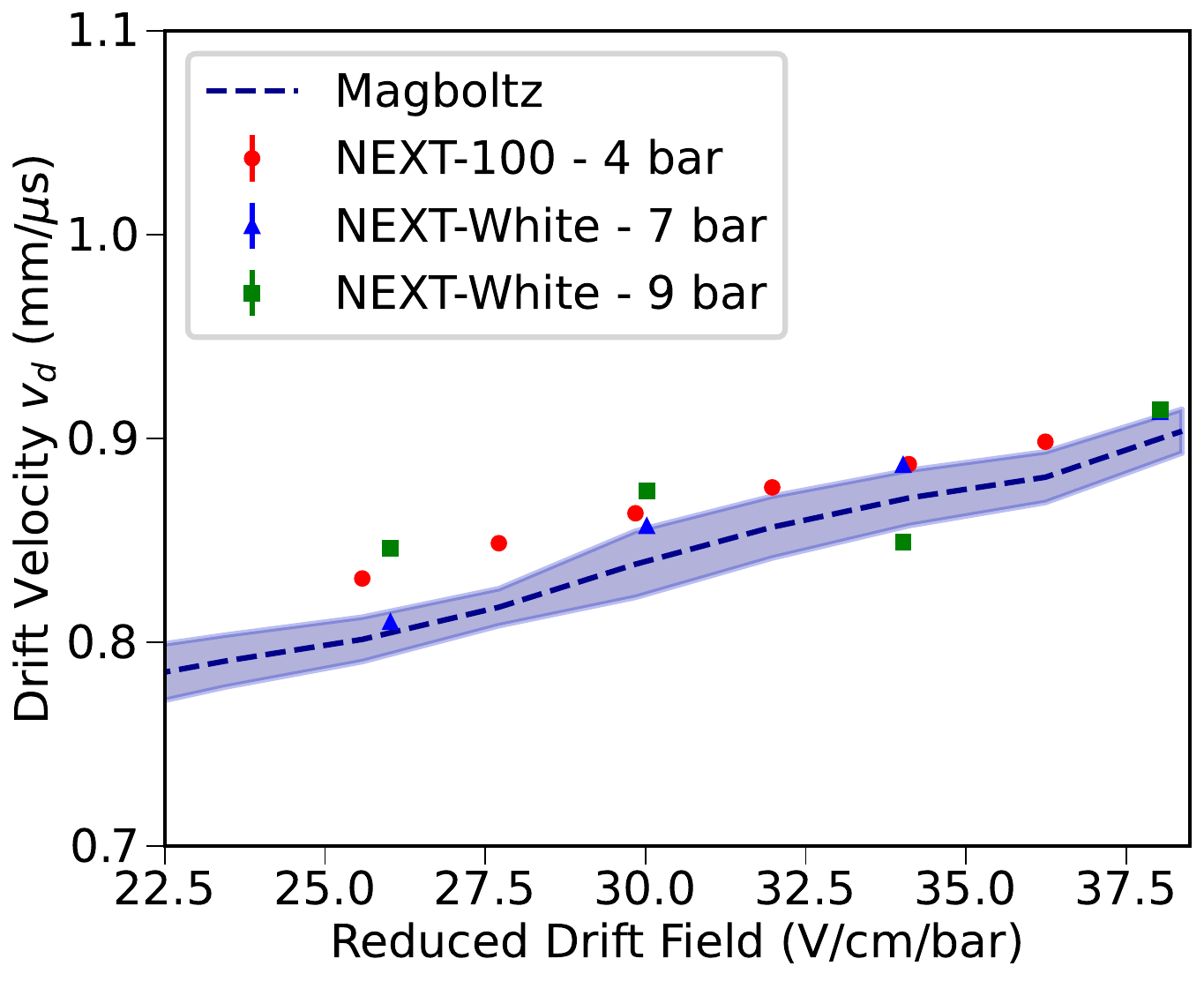}}
        \caption{}
        \label{fig:drift_vel}
    \end{subfigure}
    \begin{subfigure}{0.45\textwidth}
        \centering
        \rotatebox{0}{\includegraphics[width=1.\linewidth]{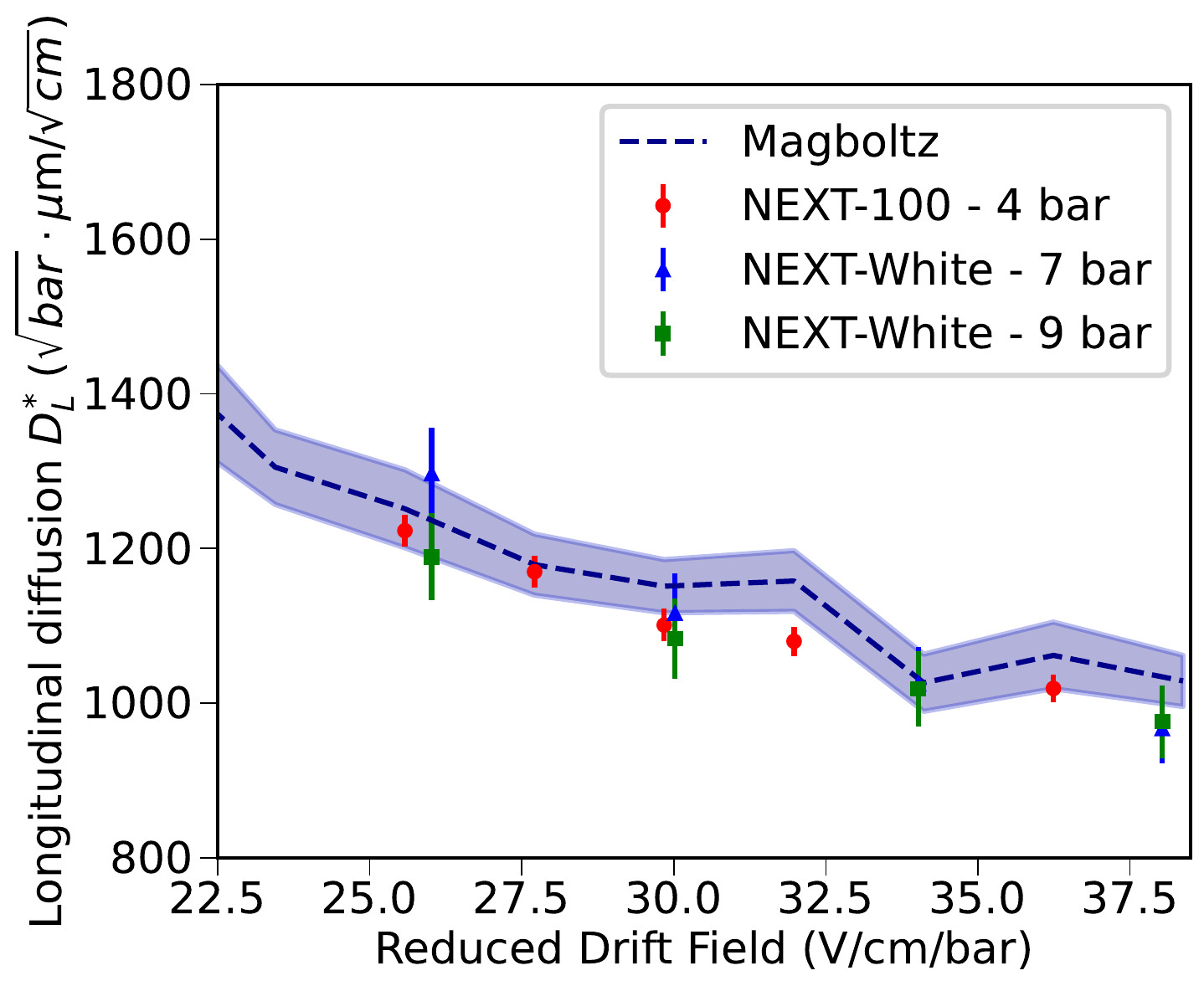}}
        \caption{}
        \label{fig:diff_long}
    \end{subfigure}
    \caption{(a) Drift velocity in gas xenon (red) for different drift conditions using reconstructed alpha events. (b) Longitudinal diffusion coefficients in gas xenon (red) for different drift conditions using reconstructed alpha events. For both plots, results are compared with Magboltz predictions (dashed blue line with 1$\sigma$ band) \cite{Magboltz}, and NEXT-White results using $^{83\text{m}}$Kr at different pressures (in green and blue) from \cite{ElectronDrift}. Uncertainties are statistical and systematic, taking into account different fit methods. }
    \label{fig:}
\end{figure}

\subsection{Commissioning with xenon (Oct. 2024 -- Feb. 2025)}

After ensuring the long-term stability of the NEXT-100 operation and the absence of gas leaks, the commissioning with xenon gas began. Following a vacuum cycle, the pressure vessel was filled with $^{136}$Xe-depleted at $\sim$\qty{4}{\bar}. As for the argon run, the operation conditions were defined via a scan of the drift fields and EL amplification fields. The voltage at the cathode ranged from \qtyrange{17}{23}{\kilo\volt}, while keeping the voltage at the gate at \qty{6}{\kilo\volt}. This corresponds to a drift field scan between {$\sim$\qty{103}{\volt/\centi\meter}} and $\sim$\qty{146}{\volt/\centi\meter} (drift velocities $\sim$\qtyrange{0.8}{0.87}{\milli\meter/\micro\second}). For a fixed drift field, an EL scan was performed between 6.5 kV and 9.0 kV, yielding from \qtyrange{1.6}{2.3}{\kilo\volt/\centi\meter/\bar}. The final voltages for this period were set at \qty{23}{\kilo\volt} for the cathode and \qty{9}{\kilo\volt} for the gate (see Tab.~\ref{table:run_pars}). These values were higher than the argon ones, but still limited by the breakdown voltage of the gate ($\sim$\qty{9.2}{\kilo\volt}) \cite{Breakdown}.

With an S2-based trigger, similar to the one used in argon data, daily samples of alpha decays were analyzed in order to monitor the evolution of the electron lifetime, the light yield, and electron drift properties. The reconstructed energy of the alpha decays of $^{222}$Rn (\qty{5.59}{\mega\electronvolt}), $^{218}$Po (\qty{6.11}{\mega\electronvolt}) and $^{214}$Po (\qty{7.83}{\mega\electronvolt}) is shown in Fig.~\ref{fig:alpha_energy}. The electron lifetime, extracted from the energy dependence with the drift time, was found to be long enough to comfortably meet specifications required for an energy resolution $\le1\%$~FWHM at the $Q_{\beta\beta}$. The attachment effect is small, making it difficult to quantify it precisely. Unlike in the NEXT-White detector, we have not found significant spatial variations of the electron lifetime. Future Krypton calibration analyses will address this question in more detail. As an example, Fig.~\ref{fig:lifetime} shows the fit of a 24-hour data sample with fluctuations typically due to the temperature variations in the LSC. The light yield was also stable below 1\% within a 24-hour run, although larger ($<$10\%) variations have been observed on a weekly basis, similar to the ones reported for the NEXT-White detector \cite{NEW0nubb}. 
Additionally, alpha decays were studied under different field conditions to evaluate the drift velocities and diffusion parameters in the chamber. The drift velocity ($v_d$) was estimated by measuring the drift time of point-like alpha events on the cathode, 
\begin{equation}
    v_d = \frac{Z_{cath}}{t_{drift} - t_{EL/2}}\quad . 
\end{equation}
This calculation uses the cathode plane as reference point in the chamber (using $Z_{cath}$ as the drift length on Tab.~\ref{table:run_pars}), and extracts drift time ($t_{drift}$) from the end-point of the drift time distribution. This relation also takes into account that the electron drift time is defined as the time difference between the S1 and S2
signals, with the S2 peak being the time when the center of the electron cloud passes the center of the EL region. This delay time {$t_{EL/2} = $\SI{0.83(0.03)}{\micro\second}} has been estimated using Magboltz \cite{Magboltz} considering the amplification voltage and the operation conditions. The different drift velocity values obtained between 103-\qty{146}{\volt/\centi\meter} are represented in Fig.~\ref{fig:drift_vel}. They show consistent results with respect to Magboltz simulations and previous results from the NEXT-White detector \cite{ElectronDrift}. The longitudinal diffusion coefficient ($D_L$) was obtained by studying the distribution of the pulse width ($\sigma^{2}_{t_{drift}}$) as a function of $t_{drift}$ \cite{ElectronDrift}
\begin{equation}
    \sigma^{2}_{t_{drift}} = \sigma^{2}_{t_{drift,0}}+2\frac{D_{L}}{v_{d}}t_{drift}\quad, 
\end{equation}
setting $\sigma^{2}_{t_{drift,0}}$ to the pulse variance at zero drift time. In this estimation, the detector is divided into drift time intervals, and the mean S2 waveform of each interval is computed. The diffusion parameters were converted to the reduced coefficient representation
\begin{equation}
    D^{*}_{L}=\sqrt{\frac{T_0}{T}\frac{2\cdot D_{L}\cdot P}{v_d}}
\end{equation}
where $T_0$ is the reference temperature (293.15 K), $T$ is the detector temperature, $P$ is the detector pressure, $D_L$ is the longitudinal diffusion coefficient, and $v_d$ is the previously estimated drift velocity value. The values obtained for $D^{*}_{L}$ are shown in Fig.~\ref{fig:diff_long}, and are compatible within uncertainties with Magboltz predictions and with previous studies from NEXT-White \cite{ElectronDrift}.

\begin{figure}
    \centering
    \begin{subfigure}{0.45\textwidth}
        \centering
        \rotatebox{0}{\includegraphics[width=1.0\linewidth]{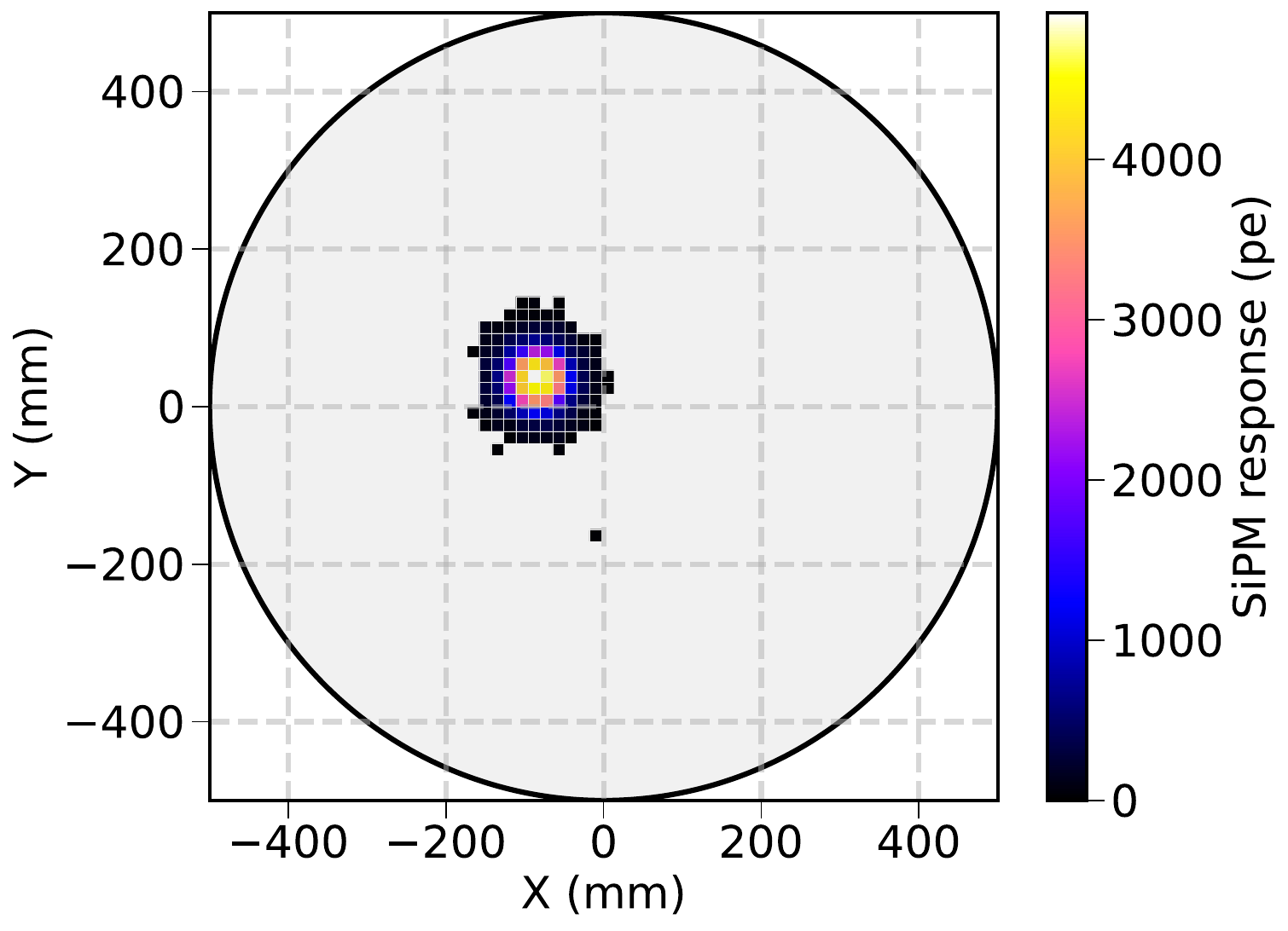}}
        \caption{}
        \label{fig:AlphaCandidate}
    \end{subfigure}
    \vspace{5mm}
    \begin{subfigure}{0.45\textwidth}
        \centering
        \rotatebox{0}{\includegraphics[width=1.0\linewidth]{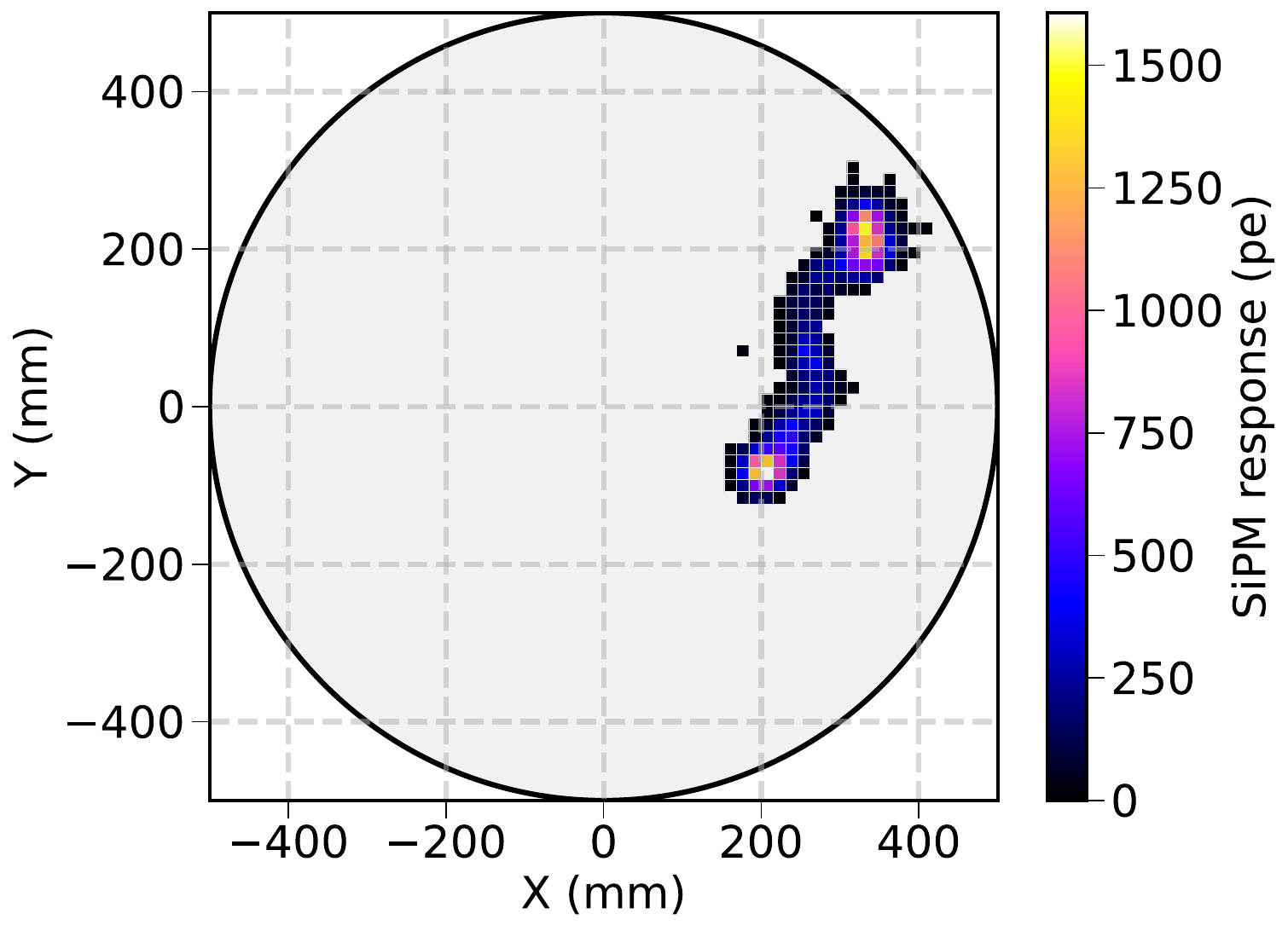}}
        \caption{}
        \label{fig:TrackCandidate}
    \end{subfigure}
    \caption{Charge collected at the SiPM plane from candidate event of (a) alpha and (b) double electron track from thallium-208 decay. Each square corresponds to a single SiPM signal. A charge threshold was applied to both event displays for visual purposes. }
    \label{fig:Commissioning}
\end{figure}

In addition, xenon data was also used to test the topological imaging capabilities of NEXT-100. As an example, Fig.~\ref{fig:AlphaCandidate} and \ref{fig:TrackCandidate} show the reconstruction of an alpha decay candidate and a double electron track, respectively. The collected charge by the tracking plane allows for distinguishing between point-like charge events (alpha decays) and extended tracks (electrons).

\subsection{Sensors Calibration}
The sensor calibration process for NEXT-100 is largely based on the one carried out on the NEXT-White detector \cite{NEWDetector}, aiming to evaluate and monitor the performance of both sensor planes during the operation of the detector. The calibration procedure relies on data collected with three different configurations: 1) sensors disconnected, 2) sensors operative, and 3) sensors operative and LED systems injecting light into the active volume. The first data sample allows for the characterization of the intrinsic electronic noise. The second sample is analyzed to measure the noise induced by the sensors and their dark count rate, identifying possible dead channels. Finally, the third sample provides the single photoelectron spectrum in each channel, which allows for the pedestal, the gain (ADC to photo-electron conversion) and the charge resolution can be derived.  

\begin{figure}
    \begin{subfigure}{0.45\textwidth}
        \centering
        \rotatebox{0}{\includegraphics[width=1.0\linewidth]{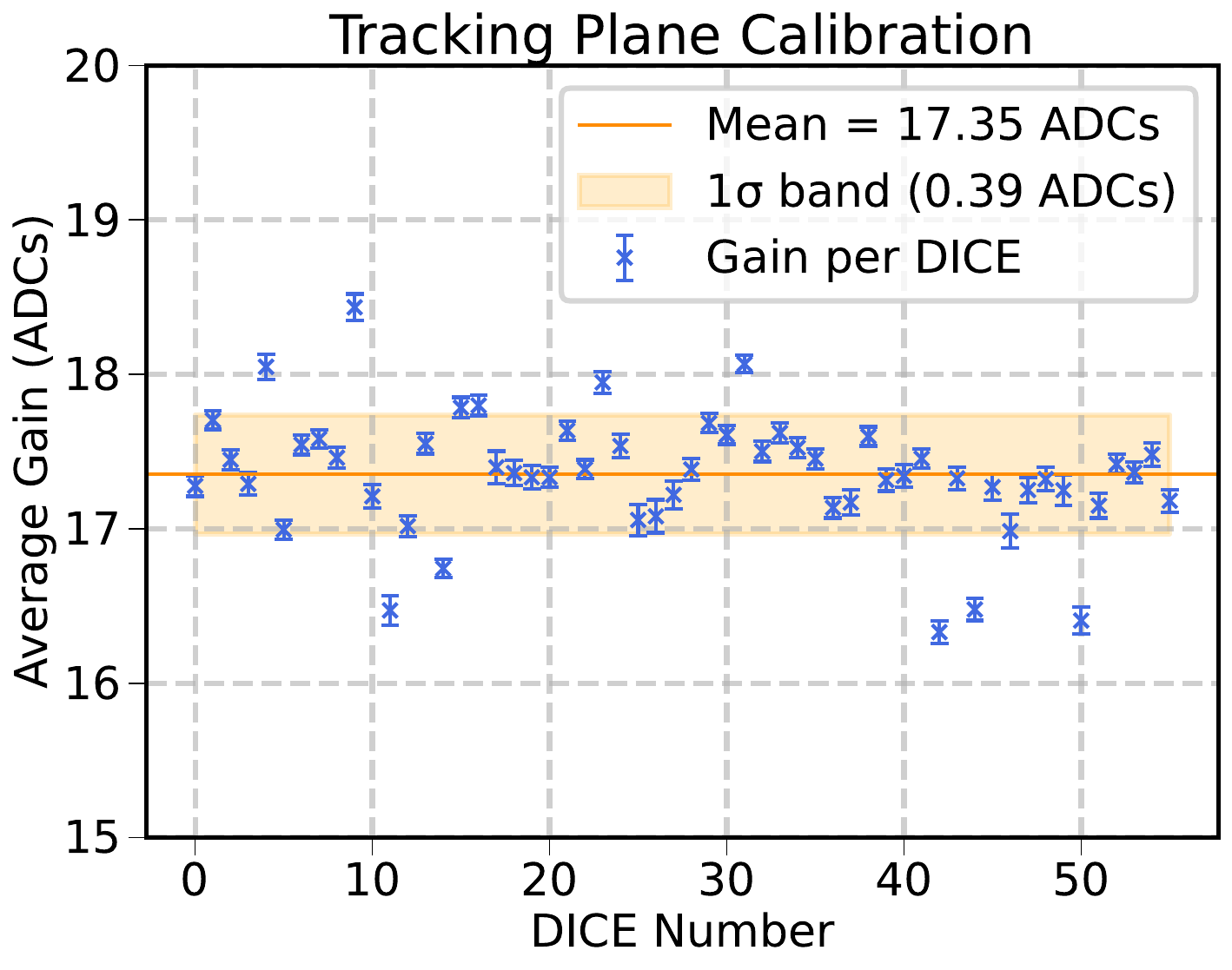}}
        \caption{}
        \label{fig:CalibStabSiPMs}
    \end{subfigure}
    \begin{subfigure}{0.45\textwidth}
        \centering
        \rotatebox{0}{\includegraphics[width=1.0\linewidth]{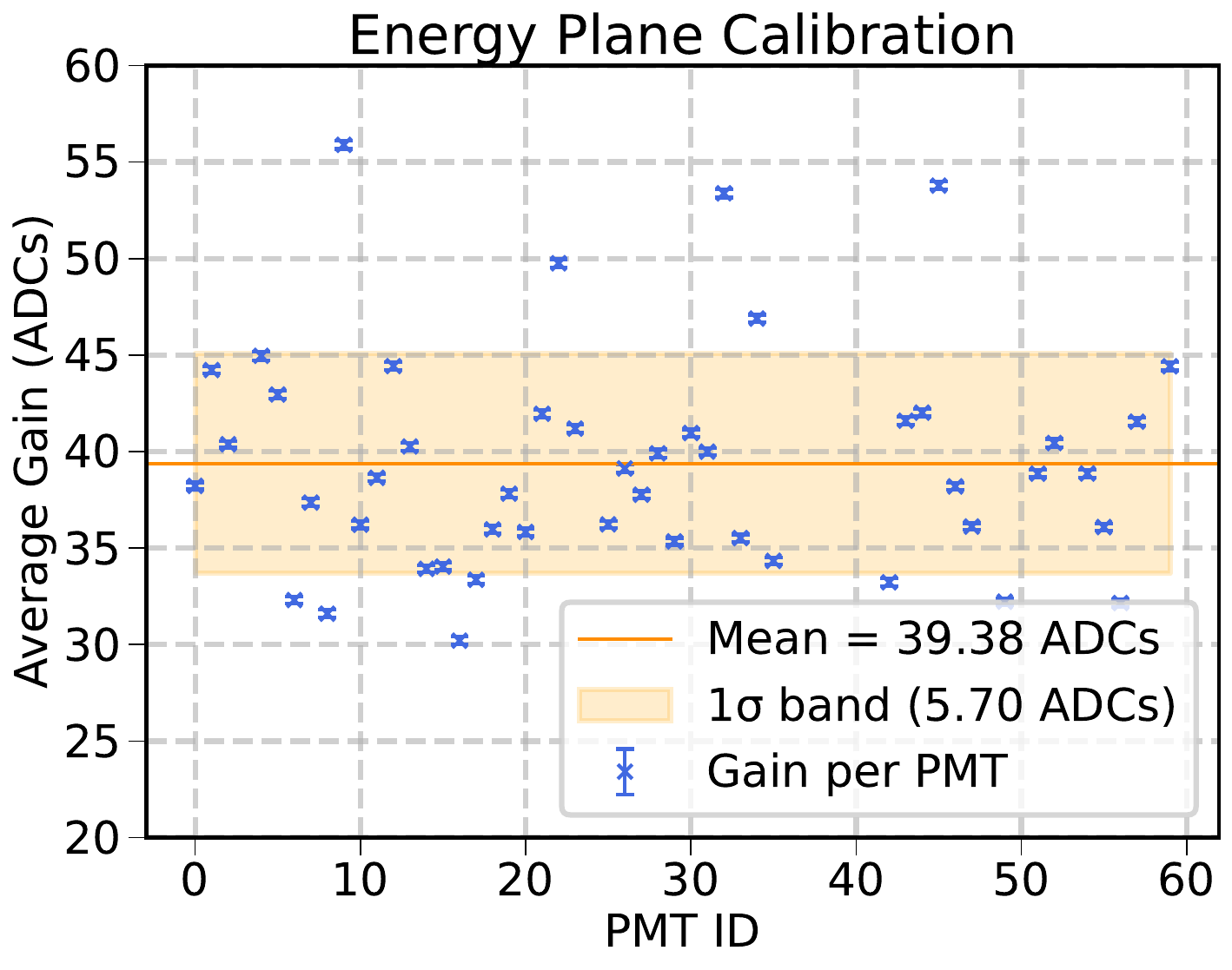}}
        \caption{}
        \label{fig:CalibStabPMTs}
    \end{subfigure}
    \caption{(a) Average gain of the DICE boards. (b) Average gain of the PMTs.}
    \label{fig:CalibStab}
\end{figure}

During the commissioning period, sensor performance and stability were monitored by taking periodic calibration runs. After the commissioning period, a dedicated calibration campaign was executed in the early months of 2025 for the evaluation of the sensor gains, which are to be used in subsequent data taking. Currently, calibration constants are monitored with weekly dedicated runs. In Fig.~\ref{fig:CalibStabSiPMs} and Fig.~\ref{fig:CalibStabPMTs}, the average gain of the SiPMs on each dice board and the gains of the PMTs are represented showing good agreement within the detector. 

\section{Summary}
The NEXT program aims to advance high-pressure xenon gas time projection chambers with electroluminescence amplification for the search for neutrinoless double beta decay. The detector NEXT-White, which operated at Laboratorio Subterr\'aneo de Canfranc from 2016 to 2021, demonstrated the robustness of this technology, the energy resolution that this type of time projection chambers can achieve, contributed to background modeling, measured the two-neutrino double beta decay mode and provided a proof-of-principle neutrinoless double beta decay analysis. 

In the next phase, the NEXT-100 experiment, began its operation in 2024 and has concluded its commissioning phase. The NEXT-100 detector is a high pressure xenon time projection chamber using electroluminescence as amplification that has proven the scalability of NEXT technology. It measures charged particles interactions with two face-to-face sensor planes. A photo-multiplier plane that enables the tagging of events and measurement of their energy; and a plane formed by silicon photo-multipliers that provides spatial and topological information of the energy depositions. The NEXT-100 experiment aims to measure the radioactive budget of the NEXT program and to improve the sensitivity to neutrinoless double beta decay searches, aiming for an energy resolution of $\sim$1$\%$ FWHM at the double beta decay energy. 

This paper gathers the information about the detector design, the assembly process, and the data acquisition improvements. Details of the improved gas system and radiopurity and material screening have also been covered. The NEXT-100 detector began taking data in May 2024 and underwent commissioning in two phases. The first phase (May–Sept. 2024) used argon gas at $\sim$\qty{4}{\bar} to test all subsystems, ensuring stability before introducing xenon in the chamber. Sensor calibration and event reconstruction were validated using alpha decays from $^{222}$Rn. In October 2024, the second phase commenced with Xe-136 depleted xenon under the same pressure conditions. Alpha events were used for initial calibration, assessing light collection and electron lifetime. The commissioning results presented in this paper have demonstrated reliable operation, paving the way for detailed physics measurements, like the study of $^{83\text{m}}$Kr events and external high energy sources. Both data samples are currently being collected and analised. 

The xenon calibration run will be followed in 2025 by a \emph{xenon background run}. After the outer shielding is closed and air is supplied from the radon abatement system, NEXT-100 will operate in an environment virtually free of airborne $^{222}$Rn, confirmed by monitoring background rate variations and $^{222}$Rn concentrations in Hall A of the LSC. In these conditions, a preliminary radiogenic background measurement will be conducted. A xenon high-pressure run will follow the background run. The plan is to fill the detector at a pressure comprised between the NEXT-White nominal pressure of \qty{10}{bar}, and the maximum certified one for the system, \qty{13.5}{bar}. A data taking plan analogous to the one presented in this paper will be carried out at higher pressure: the operation will start with an argon run in order to ensure that no leaks are present in the detector or the gas system, followed by a xenon run. This will include a calibration and background campaign as the ones included along this paper. 

The NEXT-100 experiment is currently the largest high-pressure xenon gas time projection chamber in the world. Its successful design, assembly, and commissioning has validated the robustness and scalability of this detection technology and represents a milestone in the development of rare event searches. It has provided essential operational insights relevant to the next generation of detectors, including the handling of meter-scale time projection chambers, the implementation of stable high-voltage systems under pressure conditions, and the design of efficient gas recovery systems. By addressing these technological challenges, NEXT-100 paves the way for the deployment of a larger scale, high performance detector to search for \lessnuBB~at tonne-scale and beyond.

\begin{acknowledgements}
The NEXT Collaboration acknowledges support from the following agencies and institutions: the European Research Council (ERC) under Grant Agreements No. 951281-BOLD and 101039048-GanESS; the European Union’s Framework Programme for Research and Innovation Horizon 2020 (2014–2020) under Grant Agreement No. 860881-HIDDeN; the MCIN/AEI of Spain and ERDF A way of making Europe under grants PID2021-125475NB and RTI2018-095979, and the Severo Ochoa and Mar\'ia de Maeztu Program grants CEX2023-001292-S, CEX2023-001318-M and CEX2018-000867-S; the Generalitat Valenciana of Spain under grants PROMETEO/2021/087, ASFAE/2022/028, ASFAE/2022/029, CISEJI/2023/27, and CIDEXG/2023/16;; the Department of Education of the Basque Government of Spain under the predoctoral training program non-doctoral research personnel; the Spanish la Caixa Foundation (ID 100010434) under fellowship code LCF/BQ/PI22/11910019; the Portuguese FCT under project UID/FIS/04559/2020 to fund the activities of LIBPhys-UC; the Israel Science Foundation (ISF) under grant 1223/21; the Pazy Foundation (Israel) under grants 310/22, 315/19 and 465; the US Department of Energy under contracts number DE-AC02-06CH11357 (Argonne National Laboratory), DE-AC02-07CH11359 (Fermi National Accelerator Laboratory), DE-FG02-13ER42020 (Texas A\&M), DE-SC0019054, DE-SC0019223 and DE-SC0024438 (Texas Arlington); the US National Science Foundation under award number NSF CHE 2004111; the Robert A Welch Foundation under award number Y-2031-20200401. Finally, we are grateful to the Laboratorio Subterr\'aneo de Canfranc for hosting and supporting the NEXT experiment.
\end{acknowledgements}

\bibliographystyle{unsrturl}
\bibliography{bibliography}

\end{document}